\documentclass[runningheads]{llncs}
\usepackage[T1]{fontenc}
\usepackage[ruled,noend]{algorithm2e}
\usepackage{graphicx}
\usepackage{amsmath,amssymb,amsfonts}
\usepackage{subfigure}
\usepackage{textcomp}
\usepackage{xcolor}
\usepackage{cite}
\usepackage{verbatim}
\usepackage{url}
\begin{document}
\title{Analysis of Approximate sorting in I/O model}
%
%
\author{Tianpeng Gao\inst{1,2} \and
	Jianzhong Li\inst{2,1}}
%

%
\institute{Harbin Institute of Thechnology \and
	Shenzhen Institutes of Advanced Technology, Chinese Acadamy of Sciences\\
	\email{\{gaotp,lijzh\}@hit.edu.cn}}
\maketitle              
%

\begin{abstract}
	Sorting is a fundamental component in many different applications including web indexing engines, geographic information systems, data mining and database systems. 
	Sorting problem has been wildly researched for many years and proposed a lot of efficient algorithms\cite{Knuth1998}. 
	However, the existing optimal algorithms still cause huge costs when the input data becomes very large.
	Thus, the approximate sorting for big data is considered in this paper.
	The goal of approximate sorting for big data is to generate an approximate sorted result, but using less CPU and I/O cost.
	For big data, we consider the approximate sorting in I/O model.
	
	The quality of approximate sorting results is usually measured by the distance metrics on permutation space.
	However, the existing metrics on permutation space are not available for external approximate sorting algorithms.
	Thus, we propose a new kind of metric named External metric, which ignores the errors and dislocation that happened in each I/O block.The \textit{External Spearman's footrule metric} is an example of external metric for Spearman's footrule metric.
	Furthermore, to facilitate a better evaluation of the approximate sorted result, we propose a new metric, named as \textit{errors}, which directly states the number of dislocation of the elements. Its external metric \textit{external errors} is also considered in this paper.
	
	Then, according to the rate-distortion relationship endowed by these two metrics, the lower bound of these two metrics on external approximate sorting problem with $t$ I/O operations is proved.
	We propose a k-pass external approximate sorting algorithm, named as EASORT, and prove that EASORT is asymptotically optimal.
	
	Finally, we consider the applications on approximate sorting results. An index for the result of our approximate sorting is proposed and analyze the single and range query on approximate sorted result using this index.
	Further, the sort-merge join on two relations, where one of the relations is approximate sorted or both relations are approximate sorted, are all discussed in this paper.  
\end{abstract}

\section{Introduction}
\label{sect:introduction}

Sorting is a fundamental component in many different applications including web indexing engines, geographic information systems, data mining and database systems. 
Sorting problem has been wildly researched for many years and proposed a lot of efficient algorithms\cite{Knuth1998}.  
The cost of sorting algorithm can be bounded by $O(n\log n)$ in RAM model and by $O(\frac{n}{b} \log_{\frac{m}{b}} \frac{n}{b})$ in I/O model, where $n$ is the input data size, $m$ is the main memory size and $b$ is the block size of external memory\cite{Knuth1998,Aggarwal1988}.
The I/O model, also called as external memory model, is introduced by Aggarwal, Vitter, and Shriver\cite{Aggarwal1988}, which is well introduced in Section \ref{subsec:pre_io}.
The most of existing sorting algorithms are optimal.

However, these algorithms still cause huge costs when the input data becomes very large.
For example, we performed Samplesort\cite{enwiki-samplesort} and Terasort on 1TB datasets and the computing platform is a cluster of 33 computation nodes, interconnected by the 1000Mbps ethernet.
Samplesort takes about 42 minutes and Terasort on hadoop takes more than one and half hour. When the datasets come to 1PB, Samplesort takes more than 35 days, and Terasort takes more than 90 days even on a cluster\cite{XiangYu2020}.
Thus, the cost for a complete sorting on big data is unacceptable. 
To save the cost caused by sorting, the approximate sorting problem is mainly discussed in this paper.

The approximate sorting is to generate an approximate sorted result, but using less CPU and I/O cost. 

\subsection{Related work}
Approximate sorting was first proposed by Joachim Giesen\cite{Giesen2009}. 
They studied the comparison-based approximate sorting algorithm and show how many comparisons are necessary and sufficient in order to approximately sort $n$ data items.
They also proposed an internal approximate sorting algorithm named Asort. 
Farzad Farnoud\cite{Farnoud2016} considered the problem of approximate sorting of a data stream with limited internal storage and proposed an one-pass approximate sorting algorithm. 
They analyzed the relation between the quality of approximate sorting results and the amount of available internal memory.

There are some drawbacks of existing approximate sorting analysis. 
Firstly, the existing analysis and algorithms about approximate sorting are all in RAM model, which is inappropriate.
The main reason is that the internal sorting is fast enough due to the high speed processors.
However, the I/O speed of the external disk is several orders of magnitude slower than that of internal memory.
Most of sorting for big data are external algorithm due to the limitation of internal memory capacity.
The main bottleneck of external sorting is the number of I/O operation between internal memory and external memory.  
A linear scan based on the dataset with size 1PB will take $34.7$ hours\cite{XiangYu2020}, not to mention sorting operation. 
Furthermore, the existing theory analysis and algorithms are not suitable for I/O model. 
We call the approximate sorting in I/O model as the external approximate sorting.
Thus, external approximate sorting is mainly discussed in this paper.

Secondly, the problem that comes with approximate sorting is whether the approximate sorting result can be applied in practical applications.
The previous approximate sorting algorithms\cite{Giesen2009,Farnoud2016} were only proposed for the limitation of the computing and storage resource, and none of them considered applications on approximate sorting results.
As a fundamental problem in computer system, sorting is a subroutine for many other operations, just like query, index, aggregation, join and so on.
As far as we know, the approximate sorted result can be utilized by some operations which could tolerate a certain deviation of sorting result.
For example, sort-merge join can also be solved when the data is sorted among I/O blocks, which the data in a block is larger than the data in the former I/O blocks and smaller than the data in the latter I/O blocks, and data is unsorted in each block.
There are also some studies on the applications using approximate sorted (near-sorted) sequence\cite{Disser2017,Ben2011,Athanassoulis2014,Raman2023}.
Manos et al.\cite{Athanassoulis2014} and Raman et al.\cite{Raman2023} optimize the tree index by exploit the pre-existing data ordering to offer competitive search complexity and faster index construction.
Beside, Disser et al.\cite{Disser2017} analysis the search complexity on approximate ordered sequence without index. 
In this paper, we firstly propose an tree-structure index for approximate sorted result generated by our algorithm.
Then, the single and range query on index is discussed.
We also consider the sort-merge join on two relations, where one of the relations is approximate sorted or both relations are approximate sorted.

\subsection{Measure of the Quality of Approximate Result}

The quality of approximate sorting result, also called as the distortion of approximate sorted result, is decided by the similarity between the approximate sorting result and the totally sorted data.
If the quality of approximate sorting result is higher or the distortion is smaller, the approximate sorting result is more
similar to the totally sorted result, and vice versa.

In general, the input and output of sorting algorithm is considered as permutations in permutation space $S_n$, which is well introduced in Section \ref{subsec:pre_per}. 
The distance metric on permutation space between the output of approximate sorting and the totally sorted  result is used to measure the quality of approximate result.
There are many distance metrics on permutation space used to measure the distortion of approximate result\cite{EstivillCastro1993RightIM}. 
The wildly used metrics before are mainly Kendall tau metric, Chebyshev metric and Spearman's footrule metric\cite{Farnoud2014,Farnoud2016,Giesen2009,Diaconis1977}. 

However, these metrics are not straightforward for approximate sorting results.
It is hard to intuitively understand how good or bad the approximate sorting results are from these metrics.
For example, both Spearman's foot rule metric of the sequence (8,2,3,4,5,6,7,1) and the sequence (3,4,5,2,1,7,6,8) are 14.

Furthermore, these metrics are not available for external approximate sorting algorithms. 
Since the bottleneck of performance for an external algorithm is the number of I/O operations. 
For a single I/O operation, it transforms an I/O block, $b$ continuous items, from or into the disk.
The errors and  dislocations happened in each block can be addressed in internal memory and will not affect the I/O performance.
The metric on I/O model only needs to focus on whether data items are in their appropriate block rather than the position in sequence.

Thus, in order to facilitate a better evaluation of the approximate result on external memory model, a new metric, named as \textit{errors}, is proposed, which directly states the number of dislocation of the elements in a permutation.
For example, the \textit{errors} metric of the sequence (8,2,3,4,5,6,7,1) is 2 and the sequence (3,4,5,2,1,7,6,8) is 8.
Even though the Spearman's footrule metric is the same of two sequence, the quality of first sequence is higher than the other in term of \textit{errors} metric.

In addition, we modify the mentioned metrics to fit the characteristic of I/O model.
We call this kind of metrics as External Metric, which ignores the error or displacement happened in each block to make sure those will not influence the measure of approximate result.
In this paper, the \textit{External Spearman's footrule metric} (short for \textit{ESP}) is an example of external metric for Spearman's footrule metric.
Furthermore, external metric for \textit{errors}, called as \textit{external errors} (short for \textit{EE}),is also considered in this paper.

\subsection{Contribution}
Then, the contribution of this paper is summarized as follows.
\begin{enumerate} 
	\item[-]  A new kind of metrics,\textit{External metrics}, is proposed to facilitate a better evaluation of the approximate result on external memory model.
	The \textit{External Spearman's footrule metric} (short for \textit{ESP}) is analyzed for external approximate sorting in this paper.
	\item[-] A new metric for quality of the approximate sorting result,\textit{errors}, which directly measure the dislocations in permutation, is proposed in this paper. Furthermore, the external metric for \textit{errors}, \textit{EE} is also discussed in this paper.
	\item[-] The lower bound on  \textit{EE}  and \textit{ESP} metric on external approximate sorting problem with $t$ I/O operations is proved.
	\item[-] A k-pass approximate sorting algorithm, named as EASORT, is proposed in this paper. We also prove that EASORT is asymptotically optimal in term of \textit{EE} and \textit{ESP} metric.
	\item[-] An index for the result of our approximate sorting is proposed and the I/O cost of single and range query using this index on approximate sorted result is analyzed.
	\item[-] The sort-merge join on two relations, where one of the relations is approximate sorted or both relations are approximate sorted, are all discussed in this paper.
\end{enumerate}

The following parts of the paper are organized as follows. 
First, some useful preliminary and formal definitions of new metrics are introduced in Section \ref{sec:preliminary}.
Then, the relation between the number of I/O operations, $t$ and the metric, \textit{EE}  and \textit{ESP}, are analyzed in Section \ref{sec:lower-bound}.
In Section \ref{sec:algorithm}, we propose a k-pass external approximate sorting algorithm and analyze the average distortion of algorithm results.
Section \ref{sec:apps} presents the index, query, and join operation on our approximate sorted result.
Finally, Section \ref{sec:conclusion} concludes this paper.

\section{Problem statement}\label{sec:preliminary}
\subsection{Permutation}
\label{subsec:pre_per}
For a positive integer $n$, let $[n]=\{1,\cdots,n\}$. 
A permutation is an arrangement of members in $[n]$ into a sequence. 
For example, if $n=5$, (2,3,4,1,5) is a permutation.
$S_n$ denotes the set of all permutations of $[n]$. 
For any permutation $\pi\in S_n$ and distinct $i\in [n]$, $\pi(i)$ denotes the element of rank $i$ in permutation $\pi$ and $\pi^{-1}(i)$ denotes the rank of element $i$ in  permutation $\pi$. We also denote the identify permutation by $id=[1,2,\dots,n]$. 

The input and output of approximate sorting is regarded as permutations in permutation space $S_n$.
No matter what kind of the input data $X=x_1,x_2,x_3,…,x_n$ it is, it can be mapped to a permutation $\pi\in S_n$, which satisfies that for arbitrary distinct $i,j\in[n]$, if $x_i<x_j$, then $\pi (i)<\pi (j)$. Then, $id$ is mapped to the true order of the input data. 

Through this paper, we assume that the input permutation is chosen uniformly and randomly among permutations in $S_n$. 
In the following paper, we use $\lg$ and $\ln$ as shorthand for $\log_2$ and $\log_e$, $e$ is natural constant. 
Due to the  limited length of the paper, we omit the proof process of lemmas and theorems. The full version of this paper is in \cite{Gao2022}.

The following inequalities\cite{Farnoud2014} will be used in the rest of paper,
	\begin{equation}\label{ineq:1}
		\frac{2^{nH_2(p)}}{\sqrt{8np(1-p)}} \leq \tbinom{n}{pn} \leq \frac{2^{nH_2(p)}}{\sqrt{2\pi np(1-p)}} ,
	\end{equation}
	\begin{equation}\label{ineq:2}
		\sqrt{2\pi n}(\frac{n}{e})^n \leq n! \leq \sqrt{2\pi n}(\frac{n}{e})^{n}e^{\frac{1}{(12n)}},
	\end{equation}
where $H_2(\cdot)$ is the binary entropy function, $H_2(p) = -p \lg p - (1-p) \lg (1-p)$, and we can obtain a trivial bound $H_2(p)\leq 2p\log \frac{1}{p}$ for $0<p\leq 1/2$.

\subsection{I/O model}
\label{subsec:pre_io}
I/O model, also known as external memory model, is a standard model for understanding the performance for handling large datasets in external memory\cite{Aggarwal1988}.
It assumes that the main memory can only contain $m$ data items while the remaining is kept on disk and divides into I/O blocks. Each I/O block contains $b$ continuous elements. Only the data in internal memory can be used for computation.
Data must be transferred from disk to internal memory by I/O operations before computation of the data. An I/O operation transfers a block with size of $b$ items between internal and external memory. Also, the sorting result must be reside on external memory. Those parameters also satisfy $1<b<\frac{m}{2}, m<n$.
The I/O cost is the critical complexity of external algorithms.  

\subsection{New Metric: errors}
We propose a new metric, named as \textit{errors}.
The $errors$ represents the number of dislocation elements in a permutation, which measures the quantity of the errors happened in an approximate sorting results compared with the true order.

The formal definition for $errors$ between two permutation is
\begin{equation}\label{def:err}
	d_e(\pi,\delta) = \sum_{i=1}^{n} \pi(i) \oplus \delta(i).
\end{equation}

\begin{lemma}
	\label{lemma:err}
	$d_e:S_n \times  S_n \rightarrow \mathcal{R}$ is a right-invariant pseudo-metric on permutation space $S_n$.
\end{lemma}
\begin{proof}
	At first, if $d_e$ is a pseudo-metric on permutation space $S_n$, such that for every $\pi,\delta,\theta \in S_n$, it needs to satisfy\cite{enwiki:Pseudometric}
	\begin{itemize}
		\item[1)] $d_e(\pi,\pi) = 0$.
		\item[2)] Symmetry: $d_e(\pi,\delta) = d_e(\delta,\pi)$.
		\item[3)] Triangle inequality: $d_e(\pi,\delta) \leq d_e(\pi,\theta) + d_e(\theta,\delta)$.
	\end{itemize} 
	It is obvious that the first two condition are correct for $d_e$.
	We mainly prove the third condition. 
	From the definition of $d_e$, the right of triangle inequality is equal to $\sum_{i=1}^{n}\{\pi(i) \oplus \theta(i) + \theta(i) \oplus \delta(i)\}$.
	There are three situations. First, if $\pi(i) = \delta(i) = \theta(i)$, then $\pi(i)\oplus\delta(i) = \pi(i) \oplus \theta(i) + \theta(i) \oplus \delta(i)=0$.
	Second, if $\pi(i) \neq \delta(i)$, then $\pi(i)\oplus\delta(i) \leq \pi(i) \oplus \theta(i) + \theta(i) \oplus \delta(i)$.
	Thus, triangle inequality holds for $d_\#$. 
	
	Then, we prove that $d_e$ is right-invariant, which $d_e(\pi\theta, \delta\theta)=d_e(\pi,\delta)$.
	From the definition, $d_e(\pi\theta, \delta\theta) = \sum_{i=1}^{n}\{\pi\theta(i)   \oplus \delta\theta(i)\}$.
	Since the multiplication operation for permutation $\pi$ is rearrangement the permutation $\pi$ according to another permutation $\theta$. Thus, if $\pi(i) = \delta(i)$, then $\pi\theta(i) = \delta\theta(i)$, and vice versa.
	Thus, $d_\#(\pi\theta, \delta\theta)=d_\#(\pi,\delta)$.
	\qed\end{proof}

It is clear that $0 \leq d_e\leq n$, where $d_e(\pi,\delta) = 0$ when $\pi=\delta$.
According to \cite{EstivillCastro1993RightIM}, the metric $errors$ is reasonable to measure the distortion of a permutation.

Moreover, the metric $errors$ also can be used in I/O model. In I/O model, only the dislocation happened out of each batch will be added. Thus, the formal definition of \textit{External errors}, (short for $ee$), between any two permutation $\pi,\delta\in S_n$ is 
\begin{equation}\label{def:external err}
	d_{ee}(\pi,\delta) = \sum_{i=1}^{n} \lceil\frac{\pi(i)}{b}\rceil \oplus \lceil\frac{\delta(i)}{b}\rceil.
\end{equation}
Similarly, we can prove that the external metric $d_{ee}$ is also a right-invariant pseudo-metric on permutation space $S_n$.
Furthermore, $0 \leq d_{ee}\leq n$.

\subsection{New Metric: ESP}
The formal definition of Spearman's footrule metric\cite{Diaconis1977} between any two permutation $\pi,\delta\in S_n$ is 
\begin{equation}
	d_{sp}(\pi,\delta)=\sum_{i=1}^{n}\left|\delta(i)-\pi(i)\right|\label{eq:SP definition}.
\end{equation}
This metric  is the sum of absolute values of dislocation distance between two permutation.
The maximal distance between any two permutations in $S_n$ in Spearman's footrule metric is less than $n^2$.

The definition of External Spearman’s footrule(ESP) metric between two permutation $\pi,\delta\in S_n$ is 
\begin{equation}
	d_{esp}(\pi,\delta)=\sum_{i=1}^{n}\left|\lceil\frac{\delta(i)}{b}\rceil-\lceil\frac{\pi(i)}{b}\rceil\right|\label{eq:ESP definition}.
\end{equation}
where $\lceil\frac{\delta(i)}{b}\rceil$ represents the block number of element $i$ in permutation $\delta $, and $\lceil\frac{\pi(i)}{b}\rceil$ represents the block number of element $i$ in permutation $\pi$. 
$\left|\lceil\frac{\delta(i)}{b}\rceil-\lceil\frac{\pi(i)}{b}\rceil\right|$ measures the block dislocation distance of element $i$ between two permutation. 
In fact, \textit{ESP} illustrates the sum of absolute dislocation distance between the current block number and the block number in truly ordered permutation.

For any permutation, the distance of \textit{ESP} metric to $id$ is at most $\frac{n^2}{2b}$.
This metric is obvious right invariant which means $d(\pi,\delta)=d(\pi\theta,\delta\theta)$ for any $\theta\in S_n$. 

\begin{lemma}\label{lemma:desp}
	$d_{esp}:S_n \times S_n \rightarrow \mathcal{R}$ is a right-invariant pseudo-metric on permutation space $S_n$.
\end{lemma}
\begin{proof}
	First, we prove $d_{esp}$ is a pseudo-metric on permutation space $S_n$.
	For any $\pi,\delta \in S_n$, it is obvious that $d_{esp}(\pi,\pi) = 0$ and $d_{esp}(\pi,\delta) = d_{esp}(\delta,\pi)$.
	Then, we prove $d_{esp}(\pi,\delta)\leq d_{esp}(\pi,\theta) + d_{esp}(\theta,\delta)$.
	The right of this inequality is  equal to
	\begin{align*}
		& \sum_{i=1}^{n}\left\{\left|\lceil\frac{\pi(i)}{b}\rceil-\lceil\frac{\theta(i)}{b}\rceil\right| + \left|\lceil\frac{\theta(i)}{b}\rceil-\lceil\frac{\delta(i)}{b}\rceil\right| \right\} \\
		& \geq \sum_{i=1}^{n}\left|\lceil\frac{\pi(i)}{b}\rceil-\lceil\frac{\theta(i)}{b}\rceil\ + \lceil\frac{\theta(i)}{b}\rceil-\lceil\frac{\delta(i)}{b}\rceil\right| \\
		& =  \sum_{i=1}^{n}\left|\lceil\frac{\pi(i)}{b}\rceil-\lceil\frac{\delta(i)}{b}\rceil\right| 
	\end{align*} 
	Thus,  $d_{esp}(\pi,\delta)\leq d_{esp}(\pi,\theta) + d_{esp}(\theta,\delta)$ is correct.
	
	It is clear that $d_{esp}(\pi,\delta) = d_{esp}(\pi\theta,\delta\theta)$.
	Since multiplication operation for permutation $\pi$ is rearrangement the permutation $\pi$ according to another permutation $\theta$.
	Thus, multiplication operation doesn't affect the distance between two permutations.
	\qed
\end{proof}

Moreover, the relationship between \textit{ESP} and Spearman’s footrule metric is not a simple multiple relationship.
For example, for a totally sorted sequence with $n$ data item and block size is $b$. 
Without loss of generality, we assume $b$ can divide $n$. 
Then, we reverse the data in each block and generate a new sequence. 
In this sequence, $d_{esp}$ is zero while $d_{sp}$ is $\frac{n\cdot b}{2}$. 

\subsection{Problem Definition}
Thus, the formal definition of approximate sorting in I/O model is as follows.
\begin{definition}[external approximate sorting]
	For the data sequence $X$ with $n$ data items, an external approximate sorting processes $t$ I/O operations and outputs a data sequence $Y$ with the distortion $r$.  The distortion of $Y$ is the distance between the permutation $\pi \in S_n$ and $id$, where $\pi$ is the permutation mapped by $Y$.
\end{definition}

\section{Analysis of Lower Bound}\label{sec:lower-bound}
In this section, we will show the lower bound of \textit{ESP} and \textit{EE} metric for any approximate sorting algorithms in I/O model.

To derive this bound, we use the rate-distortion theory on permutation space\cite{Farnoud2014,Farnoud2016}. 
Rate–distortion theory is a major branch of information theory which provides the theoretical foundations for lossy data compression.
It addresses the problem of determining the minimal number of bits per symbol, as measured by the rate $R$, that should be communicated over a channel, so that the source (input signal) can be approximately reconstructed at the receiver (output signal) without exceeding an expected distortion $r$\cite{rate-distortion}.

The information-theoretic lower bound\cite{Knuth1998} of compare-based sorting explains that $\lg n!$ "bits of information" are being acquired during a sorting procedure. 
However, the information gain of approximate sorting is less than the information gain for totally sorting. 
Thus, the relation between the information gain of approximate sorting and the distortion of approximate result can be characterized by the rate-distortion theory\cite{Farnoud2014,Wang2013,Klove2010} on permutation space.
Clearly, the approximate sorting can be analogized to a lossy compression problem on permutation space.

Therefore, we use rate–distortion theory to dictate an universal bound of distortion for a given rate for approximate sorting in I/O model. 
According to that, we will get the relation between distortion $r$ and the number of I/O operations, $t$.
Next, we explain the rate-distortion on permutation space endowed by $d_{ee}$ and $d_{esp}$.

\subsection{Rate–Distortion Theory in I/O model  }\label{AA}
Specifically, the information gained by approximate sorting with $t$ I/O steps can be regarded as the number of permutations generated by approximate sorting on permutation space. For any right-invariant distance metric(pseudo-metric) on permutation space, we define the ball centered at $\delta \in S_n$, with radius $r$ under any metric $d$ as 
\begin{equation*} 
	Ball(\delta,r) = \{\pi|\pi\in S_n, d(\pi,\delta) \leq r \}
\end{equation*}
$Ball(\delta,r)$ represents the permutations in $S_n$ which the distance to $\delta$ is no more than $r$. $|Ball(\delta,r)|$ is the size of ball. 
Because of right-invariant, $|Ball(\delta,r)|$ is not related to its center $\delta$. We can take $Ball(\delta,r)$  as $Ball(r)$ directly.

These balls divide the permutation space into many classes, which the distance between two permutation in each class is not large than $r$. If any approximate sorting algorithm generate an approximate result $\pi$ which satisfies $d(\pi, \delta) < r$ for any $\delta \in S_n$ including the $id$, the number of possible permutations generated by the algorithm is at least larger than the size of classes according to the pigeonhole principle.

Next, some basic definitions for rate-distortion theory on permutation space are introduced.
We consider the approximate space or code space $C$ as a subset of $S_n$ (The code space is a proper term in the field of lossy compression).  
The distance between the approximate space $C$ and any permutation $\pi\in S-N$ can be defined as 
\begin{equation*}
	d(\pi, C) = \min_{\delta \in C} \{ d(\pi,\delta)\}.
\end{equation*}

We use $\hat{A}(r)$ to denote the minimum size of the approximate space $C$ required for a worst-case distortion such that $d(\pi,C) \leq r$ for any permutation $\pi\in S_n$. 
We use $\bar{A}(r)$ denote the minimum size of the approximate space $C$ required for an average distortion $r$ under the uniform distribution on $S_n$, that is $\bar{A}(r)$ is the size of approximate space $C$ such that $\frac{1}{n!}\sum_{\pi\in S_n}{d(\pi, C)\leq r}$.

We define the \textit{rate} as $\hat{R}(r) = \frac{1}{n} \lg \hat{A}(r)$ and $\bar{R}(r) = \frac{1}{n} \lg \bar{A}(r)$. 
$\hat{R}(r) (\bar{R}(r))$ means that the information gain of each permutation in $\hat{A}(r) (\bar{A}(r))$.

According to Farzad Farnoud\cite{Farnoud2014}, two basic lower bounds for $\hat{A}(r)$ and $\bar{A}(r)$ was proposed and these lower bounds are available for any right-invariant distance on permutations. 

\begin{theorem}[Farzad Farnoud\cite{Farnoud2014},Theorem4]
	\label{theorem:universal}
	For all $n,r\in \mathcal{N}$, $n$ is denoted as the data size, 
	\begin{equation} \label{th1:eq1}
		\frac{n!}{|Ball(r)|} \leq \hat{A}(r) \leq \frac{n!}{|Ball(r)|} \cdot (1+\ln |Ball(r)|),
	\end{equation}
	\begin{equation}\label{th1:eq2}
		\frac{n!}{|Ball(r)|\cdot(r+1)} \leq \bar{A}(r) \leq \hat{A}(r),
	\end{equation}
\end{theorem}

\subsection{rate-distortion relationship for $d_{ee}$}
In this subsection, we get the rate-distortion relationship for $d_{ee}$.
First, we find the bounds on the size of $Ball(r)$ based on the  $d_{e}$ and $d_{ee}$.
\begin{lemma}\label{lemma:ee ball}
	For $0 \leq r\leq n$, 
	\begin{equation}
		\left|Ball_{e}(r)\right| \leq \tbinom{n}{r} \cdot r!,
	\end{equation}
	\begin{equation}
		\left|Ball_{ee}(r)\right| \leq \tbinom{n}{r} \cdot r! \cdot b^{n-r}.
	\end{equation}
\end{lemma}

\begin{proof}
	For any $\pi \in S_n$, $\left|B_{e}(r)\right|=\left|B_{e}(\pi,r)\right| = \left|\left\{\delta|\delta \in S_n, d_{\#e}(\pi,\delta)\leq r\right\}\right|$.
	For any $\delta \in B_{e}(\pi,r)$, there are at most $r$ elements in $\pi$ is misplaced compared with $\delta$. Thus, it has at most $\tbinom{n}{r} \cdot r!$ possible permutations.
	$\tbinom{n}{r}$ represents the number of possible $r$ elements that is misplaced.
	Since $B_{e}(r)$ represents the number of dislocation elements is less than $r$, these possible $r$ elements can be arbitrary arrangement. Thus,  $\left|B_{e}(r)\right|\leq \tbinom{n}{r} \cdot r!$.
	
	For $B_{ee}(r)$, it is same as the $B_{e}$ to satisfy the distance $r$.
	But, the dislocation happened in each block is not counted for  $B_{ee}(r)$.
	For the rest $n-r$ elements, each element may have at most $b$ positions.
	Thus, $\left|B_{ee}(r)\right| \leq \tbinom{n}{r} \cdot r! \cdot b^{n-r}$.
	\qed\end{proof}

From Theorem \ref{theorem:universal},  we can get the rate-distortion relation under $d_{ee}$.

\begin{lemma}
	\label{lemma:rate-distortion for errors}
	$\forall n, r\in \mathcal{N}, 0 \leq r\leq n, \alpha=\frac{r}{n}, 0\leq \alpha \leq 1$, 
	\begin{equation}
		\bar{R}(r) \geq \frac{\lg n!}{n} -\frac{(n-r)\lg b}{n} - \alpha \lg n-\lg (r+1),
	\end{equation}
	where $n$ represents the data size and $r$ is the $d_{ee}$. 
\end{lemma} 

\begin{proof}
	For the average case distortion $r$, we have 
	\begin{equation*}
		\lg \left|Ball_{ee}(r)\right| \leq \lg \left(\tbinom{n}{r} \cdot r! \cdot b^{n-r}\right) = \lg \tbinom{n}{r} + \lg r! + (n-r)\lg b,
	\end{equation*}
	Then, simplifying it by the inequality \eqref{ineq:1}, we have
	\begin{align*}
		\lg \left|B_{ee}(r)\right| \leq n\alpha \lg n + (n-r)\lg b 
	\end{align*}
	From Theorem \ref{theorem:universal}, 
	\begin{align*}
		\bar{R}(r) = \frac{1}{n} \lg \bar{A}(r) &\geq \frac{1}{n}\left( \lg n! - \lg \left|B_{ee}(r)\right|-\lg (r+1)\right) \\ &\geq \frac{\lg n!}{n} -\frac{(n-r)\lg b}{n} - \alpha \lg n-\frac{\lg (r+1)}{n}.
	\end{align*}
	\qed\end{proof}

\subsection{rate-distortion relationship for $d_{esp}$}
The goal of this section is to get the rate-distortion relationship endowed by ESP metric. 
First, we find the bounds on the size of ball based on the ESP metric .
\begin{lemma}
	\label{lemma:esp ball}
	\begin{equation}
		\left|Ball_{esp}(r)\right| \leq (2\cdot b)^n\cdot \tbinom{n+r}{r}
	\end{equation}
\end{lemma}

\begin{proof}
	For any $\pi \in S_n$, $|Ball(r)| = \sum_{i=0}^r|\{\delta|\delta\in S_n,d(\pi,\delta) = i\}|$.
	For each non-negative integer $i$, it can be regard as the sum of $n$ non-negative integers sequence, $i_1,i_2,\dots,i_n$.
	We can define 
	\begin{equation}
		d_{esp}(\pi,\delta)=\sum_{j=1}^n \left|\lceil\frac{\pi(j)}{b}\rceil -\lceil\frac{\delta(j)}{b}\rceil \right| = \sum_{j=1}^n i_j = i 
	\end{equation}
	The possible unsigned solutions of equation $\sum_{j=1}^n i_j = i$ is equal to the number of ways of placing $i$ balls into $b$ bins, which is equal to $\tbinom{n+i-1}{i}$.
	For any solution sequence $i_1,i_2,\dots,i_n$, each $i_j $is related to at most $2\cdot b$  permutations.
	Therefore, for any $i$, 
	\begin{equation*}
		\left\{ \delta|\delta \in S_n, d_{esp}(\pi,\delta) = i\right\} \leq (2\cdot b)^n\cdot\tbinom{n+i-1}{i},
	\end{equation*} and hence
	\begin{equation*}
		\left|Ball_{esp}(r)\right| \leq \sum_{i=0}^r{(2\cdot b)^n\cdot\tbinom{n+i-1}{i}}=(2\cdot b)^n\cdot\tbinom{n+r}{r}.
	\end{equation*}
	\qed\end{proof}

According to Lemma \ref{lemma:esp ball} and Theorem \ref{theorem:universal}, we can get rate-distortion relation under ESP metric. 

\begin{lemma}
	\label{lemma:rate-distortion esp}
	For all $n,r\in N$,$n$ is denoted as the data size and $r$ is the ESP distance, $\alpha=\frac{r}{n}, 0\leq \alpha \leq \frac{n}{2b}$.
	\begin{equation}\label{eq:rate1}
		\bar{R}(r) \geq \frac{\lg n!}{n} - \frac{\lg 2b}{n} - \lg {\frac{(1+\alpha)^{1+\alpha}}{\alpha^\alpha}} - \frac{\lg n}{n}.
	\end{equation}
\end{lemma}
\begin{proof}
	For the average-case distortion $r$, we have 
	\begin{equation*}
		\lg {|Ball_{esp}(r)| \cdot (r+1)} \leq \lg{(2\cdot b)^n} + \lg{\tbinom{n+r}{r}\cdot (r+1)},
	\end{equation*}
	From  inequality \eqref{ineq:1}  , we have 
	\begin{align*}
		\lg{\tbinom{n+r}{r}\cdot (r+1)} &= \lg{\tbinom{n(1+\alpha)}{n\alpha}}\cdot (n\alpha+1) \\ 
		& \leq \lg \left\{{2^{n(1+\alpha)H(\frac{1}{1+\alpha})}}\cdot \frac{n\alpha +1 }{\sqrt{{2\pi n\alpha}/(1+\alpha)}}\right\} \\
		& = \lg\left\{{2^{n(1+\alpha)H(\frac{1}{1+\alpha})}} \cdot \sqrt{\frac{2\alpha n}{\pi}} \cdot \frac{1+\frac{1}{n\alpha}}{2\sqrt{n\alpha (1+\alpha)}}\right\},
	\end{align*} 
	because the expression $\frac{1+\frac{1}{\alpha n}}{2\sqrt{n\alpha (1+\alpha)}}$ is decreasing when $\alpha$ is increasing and it get maximized when $\alpha=\frac{1}{n}$. 
	Hence, 
	\begin{equation*}
		\frac{1+\frac{1}{\alpha n}}{2\sqrt{n\alpha (1+\alpha)}} \leq \frac{1}{\sqrt{1+\frac{1}{n}}}\leq 1.
	\end{equation*}
	So, according to the above formula and inequality, 
	\begin{equation*}
		\lg{\tbinom{n+r}{r}\cdot (r+1)} \leq \lg\left\{{2^{n(1+\alpha)H(\frac{1}{1+\alpha})}} \cdot \sqrt{\frac{2\alpha n}{\pi}}\right\},
	\end{equation*}
	From Theorem \ref{theorem:universal},
	\begin{align*}
		\bar{R}(r) = \frac{1}{n} \lg \bar{A}(r) &\geq \frac{1}{n}\left( \lg n! - \lg \left|B_{esp}(r)\right|\cdot(n\alpha+1)\right)\\ &\geq \frac{\lg n!}{n} -\frac{2b}{n}-\lg \frac{(1+\alpha)^{1+\alpha}}{\alpha^\alpha} - \frac{\lg n}{n}. 
	\end{align*}
	\qed\end{proof}

Next, we will use the rate distortion theory in Lemma \ref{lemma:rate-distortion for errors} and \ref{lemma:rate-distortion esp} to find out the relation between the number of I/O operations $t$ and distortion $r$ of approximate results. 
We will provide the lower bound on \textit{EE} and \textit{ESP} metric respectively between $id$ and the approximate sorting results generated in $t$ I/O operations. 

We denote the distance as $r$, and the $\alpha$ denotes the average errors or dislocation distance of each elements.

\begin{lemma}
	\label{lemma:code size}
	For an external approximate sorting algorithm $f$, the upper bound of $\bar{A}(r)$ after processing $t$ I/O operations is 
	\begin{equation}
		\bar{A}(r) \leq \frac{n}{b}^t\cdot \tbinom{m}{b}^{\frac{t}{2}}\cdot (b!)^{min(\frac{n}{b},\frac{t}{2})}.
	\end{equation}
\end{lemma}

\begin{proof}
	From the definition of $\bar{A}(r)$, $\bar{A}(r)$ in a specific external approximate sorting means the number of different possible permutations generated by the algorithm.
	First, we find out the upper bound for the size of $\bar{A}(r)$ after $t$ I/O operations.
	According to the introduction of I/O model, it is clear that the number of read operation and write operation of an approximate sorting algorithm is equal, which is $t/2$. 
	
	A read operation specifies which block out of $\frac{n}{b}$ nonempty blocks is to be read. 
	A write specifies which elements are to be written and in which empty blocks. 
	There are at most $\tbinom{m}{b}$ choices for a write operation to choose different elements in memory to be written. 
	Each block which have not reside in internal memory before will increase the possible permutation by factor $b!$. 
	This condition only happen once for each $\frac{n}{b}$ block.
	Hence, we have
	\begin{equation*}
		\bar{A}(r) \leq \frac{n}{b}^t\cdot \tbinom{m}{b}^{\frac{t}{2}}\cdot (b!)^{min(\frac{n}{b},\frac{t}{2})}.
	\end{equation*}
	\qed\end{proof}

\subsection{The Lower Bound of $d_{ee}$}
\begin{theorem}
	\label{theorem:external error}
	For any external approximate sorting algorithm $f$ take $t$ I/O operations, the expected \textit{EE} of permutation generated by algorithm to the totally sorted permutation $id$ is r. Let $\alpha = \frac{r}{n}$. If $0 \leq \alpha \leq \frac{n-1}{n}$, then
	\begin{equation}
		\alpha \geq  \log_{\frac{n}{b}}{\frac{nb^{(\frac{t}{2}-\min(\frac{n}{b},\frac{t}{2}))\frac{b}{n}}}{2e\cdot b^2\cdot\frac{n}{b}^{\frac{t}{n}}\cdot (em)^{\frac{t}{2}\cdot \frac{b}{n}}}}.
	\end{equation}
\end{theorem}
	\begin{proof}
		From Lemma \ref{lemma:code size}, we have
		\begin{align*}
			\lg {\bar{A}(r)} &\leq \lg  {(\frac{n}{b})^t\cdot \tbinom{m}{b}^{\frac{t}{2}}\cdot (b!)^{\min(\frac{n}{b},\frac{t}{2})}} \\
			&= t\cdot \lg \frac{n}{b} + \frac{t}{2}\cdot \lg \tbinom{m}{b}+ \min(\frac{n}{b},\frac{t}{2})\cdot \lg b!.
		\end{align*}
		Simplify above formula, we get that 
		\begin{align*}
			\lg {\bar{A}(r)} &\leq t\cdot \lg \frac{n}{b} + \frac{t}{2}\cdot b\cdot \lg \frac{em}{b} + \min(\frac{n}{b},\frac{t}{2})\cdot b\cdot \lg b \\
			&= t\cdot \lg \frac{n}{b} + \frac{t}{2}\cdot b\cdot \lg  {em}+\left(\min(\frac{n}{b},\frac{t}{2})-\frac{t}{2}\right)\cdot b \cdot \lg b.
		\end{align*}
		Thus,
		\begin{equation}
			\label{ineq:rate}
			\bar{R}(r) = \frac{1}{n}\cdot \lg \bar{A}(r) \leq \frac{t}{n}\cdot \lg \frac{n}{b}+\frac{t}{2}\cdot \frac{b}{n}\cdot \lg {em} + \left(\min(\frac{n}{b},\frac{t}{2})-\frac{t}{2}\right)\cdot \frac{b}{n}\cdot \lg b.
		\end{equation}
		Based on Lemma \ref{lemma:rate-distortion for errors} and above equation, we have
		\begin{align*}
			\alpha \lg n + \frac{\lg (n\alpha + 1)}{n} -\alpha \lg b &\geq \frac{\lg n!}{n} - 2\lg b - \frac{t}{n} \lg \frac{n}{b}-\frac{t}{2} \frac{b}{n} \lg {em} \\ &-\left(\min(\frac{n}{b},\frac{t}{2})-\frac{t}{2}\right)\frac{b}{n} \lg b.
		\end{align*}
		Simplify above equation, we have
		\begin{equation*}
			(\frac{n}{b})^\alpha \geq \frac{nb^{(\frac{t}{2}-\min(\frac{n}{b},\frac{t}{2}))\frac{b}{n}}}{2e\cdot b^2\cdot\frac{n}{b}^{\frac{t}{n}}\cdot (em)^{\frac{t}{2}\cdot \frac{b}{n}}}
		\end{equation*}
		Thus, 
		\begin{equation}
			\alpha \geq  \log_{\frac{n}{b}}{\frac{nb^{(\frac{t}{2}-\min(\frac{n}{b},\frac{t}{2}))\frac{b}{n}}}{2e\cdot b^2\cdot\frac{n}{b}^{\frac{t}{n}}\cdot (em)^{\frac{t}{2}\cdot \frac{b}{n}}}}.
		\end{equation}
		\qed\end{proof}

Next, we analyze the asymptotic bound for different regimes of $t$. 
It is worthy noting that the linear time on I/O model is $O(\frac{n}{b})$ rather than $O(n)$.
Moreover, $\alpha$ is represent the average external errors for each data element.
According to Theorem \ref{theorem:external error}, we have 
\begin{enumerate}
	\item[-] When $t=O(\frac{n}{b}\cdot \log_{\frac{m}{b}} \frac{n}{b})$, $\left((\frac{n}{b})^{\frac{t}{n}}\cdot (em)^{\frac{t}{2}\cdot \frac{b}{n}}\right)$ is close to $O(n)$. Then, $\alpha > 0$.
	\item[-] When $t=2k\cdot\frac{n}{b}$, $\left((\frac{n}{b})^{\frac{t}{n}}\cdot (em)^{\frac{t}{2}\cdot \frac{b}{n}}\right)$ is close to $O((em)^{k})$.Let $\frac{n}{b}= O(\frac{m}{b}^c), c>k$, then $\alpha >1-\frac{k}{c}$.
	\item[-] When $t=o(\frac{n}{b})$, $\left((\frac{n}{b})^{\frac{t}{n}}\cdot (em)^{\frac{t}{2}\cdot \frac{b}{n}}\right)$ is close to $n$. $\alpha$ is close to $1$.
\end{enumerate}

\subsection{The Lower Bound of $d_{esp}$}
\begin{theorem}
	\label{theorem:esp}
	For any external approximate sorting algorithm $f$ take $t$ I/O operations, the expected ESP metric of algorithm output permutation to totally sorted permutation $id$ is $r$, we have
	\begin{equation}
		r \geq \frac{n^2b^{(\frac{t}{2}-\min(\frac{n}{b},\frac{t}{2}))\frac{b}{n}}}{e^2\cdot 2b\cdot (\frac{n}{b})^{\frac{t}{n}}\cdot (em)^{\frac{t}{2}\cdot \frac{b}{n}}}-n
	\end{equation}
\end{theorem}
	\begin{proof}
		The proof of relation between $esp$ and  the number of I/O operation $t$ is similar to Theorem \ref{theorem:external error}.
		
		From the inequality \eqref{ineq:rate} and Lemma \ref{lemma:rate-distortion esp}, we have
		\begin{align*}
			\lg \frac{(1+\alpha)^{1+\alpha}}{\alpha^\alpha} &\geq \frac{\lg n!}{n} - \frac{\min(n,\delta n)}{n}\cdot \lg 2b - \frac{\lg n}{n} - \frac{t}{n}\cdot \lg{\frac{n}{b}\cdot (em)^{\frac{t}{2}}} \\ &-\left(\min(\frac{n}{b},\frac{t}{2})-\frac{t}{2}\right)\frac{b}{n} \lg b,
		\end{align*}
		Then, the left side of above formula can be simplified as 
		\begin{equation*}
			\lg \frac{(1+\alpha)^{1+\alpha}}{\alpha^\alpha} = \lg {(1+\frac{1}{\alpha})^\alpha \cdot (1+\alpha)}\leq \lg {e(1+\alpha)},
		\end{equation*}        
		Next, we simplify the right side of equation and combined with above formula.
		\begin{equation*}
			\alpha \geq \frac{nb^{(\frac{t}{2}-\min(\frac{n}{b},\frac{t}{2}))\frac{b}{n}}}{e^2\cdot 2b\cdot (\frac{n}{b})^{\frac{t}{n}}\cdot (em)^{\frac{t}{2}\cdot \frac{b}{n}}}-1
		\end{equation*}
		Hence, we have 
		\begin{equation*}
			r \geq \frac{n^2b^{(\frac{t}{2}-\min(\frac{n}{b},\frac{t}{2}))\frac{b}{n}}}{e^2\cdot 2b\cdot (\frac{n}{b})^{\frac{t}{n}}\cdot (em)^{\frac{t}{2}\cdot \frac{b}{n}}}-n, 
		\end{equation*}
		\qed\end{proof}

Next, we discuss the asymptotic bound for different regimes of I/O steps.
At first, we need know that linear time on external model is $O(\frac{n}{b})$ rather than $O(n)$. 
Moreover, we use $\alpha=\frac{r}{n}$ to denote the average block deviation on single data item.
According to above theorem, we have
\begin{enumerate}
	\item[-] When $t=O(\frac{n}{b}\cdot \log_{\frac{m}{b}} \frac{n}{b})$,  then we have $\alpha > o(1)$.
	\item[-] When $t=2k\cdot \frac{n}{b}$, then we have $\alpha > O\left( \frac{nb^{k-2}}{ (em)^{k}}\right)-1$.
	\item[-] when $t=o(\frac{n}{b})$,  then we have $\alpha > O\left(\frac{n}{b\cdot (em)^{o(1)}}\right)-1 $ 
\end{enumerate}
we know that the max block deviation of each item is $\frac{n}{b}$.
$\alpha$ decrease by a factor $em$ with every $2\cdot\frac{n}{b}$ I/O steps.

\section{Approximate Algorithm}\label{sec:algorithm}
\subsection{Design of External Approximate Sorting Algorithm}
\begin{algorithm}[h]
	\small
	\caption{EASORT}
	\label{alg:1}
	\LinesNumbered
	\KwIn{	{\bf $\frac{n}{b}$ blocks, the parameter $k$ and internal memory size $m$}} 
	\KwOut{	{\bf An approximate sorting result on external memory} }
	
	Compute the number of buckets, $p = \lfloor\frac{(m-b)}{(b+1)}\rfloor$\;
	$i = 0$\;
	\If{$i<k$}{
		\For {each bucket: B}{
			\If{$|B|< m$}{
				read all items in $f$ into internal memory and sort them\;
				\textbf{break}\;
			}
			Pick random $\frac{m}{b}$ blocks into memory and sort them\;
			Choose pivots $P_1,\cdots,P_{p-1}$\;
			Set buffers: $\mathcal{B}_1, \cdots, \mathcal{B}_p = \emptyset$\;
			\For {each data $e$ in each block} {
				\If{$P_i<e<P_{i+1}$}{
					$\mathcal{B}_i = \mathcal{B}_i \cup \{e\}$\;
					\If{$|\mathcal{B}_i| == b$}{
						Sort($\mathcal{B}_i$) and  write to corresponding bucket file\;
						$\mathcal{B}_i = \emptyset$\;
					}
				}
			}			
			Write all data in each buffer into their corresponding file\;
			Remove the original data file\;
		}
		$i++$\;
	}
	\Return ALL bucket files.
\end{algorithm}
In this section, we present a k-pass I/O approximate sorting algorithm. 
One pass I/O means that each data block are only read and written once and the total number of  I/O operations is $t=2\cdot \frac{n}{b}$.

For sorting problem on external memory, algorithms are mainly divided into two types\cite{Aggarwal1988}, merge-based sorting algorithms, such as multi-way merge sort and distribution-based sorting algorithms, such as sample sort. Those have been researched for a long time and they have proposed a lot of algorithms for different practical application. 

The Merge-based sorting algorithm consists of two phase. 
The first phase is the ‘run formation’, which is to read a large enough data block into internal memory and sort them to form an ordered sequence called ‘run’ until all data is processed once. Then, the second phase is to recursively merge all the runs together. 

The Distribution-based sorting algorithm first divides the data into different bucket according to some methods, which keep the data between different buckets into an ascending order.
Then, it sorts the data in each buckets.

For approximate sorting, it tries to keep all data as close as possible to their right blocks.
Thus, we proposed an external approximate sorting algorithm in a distribution way.
The intuitive idea is that we divide all data into many buckets and put as much data as possible into the correct buckets. The bucket size is related to the distortion of approximate result.
If the bucket size is equal to the I/O block size, then the distortion of this approximate sorting is zero in I/O model. Since all data are in their right blocks.

The k-pass algorithm EASORT is shown in Algorithm \ref{alg:1}.

As shown in Algorithm \ref{alg:1},  the main idea is to divide the original data into many buckets in each pass.
Firstly, it needs to determine the number of buckets, $p$.
Each bucket needs at least on internal buffer, whose size is equal to the I/O block.
Only then can ensure that each write operation of algorithm can write at least one I/O block to external memory.
Thus, $p\cdot b+b+p\leq m$, and $1\leq p\leq \lfloor\frac{(m-b)}{(b+1)}\rfloor$ (line 1).

After computing $p$, it checks whether the size of bucket file is smaller than the size of internal memory(lines 4-5).
If it is, then Algorithm \ref{alg:1} use the internal algorithm to sort it and output the result(lines 6-7).
If not, it reads at most $m$ items into internal memory and choose $p-1$ elements as pivots which can divide such $m$ items into $p$ buckets evenly (lines 8-9).
These pivots are sorted in ascending order. 
Assume that sorted $p-1$ pivots are $P_1, \cdots, P_{p-1}$ and $P_0,P_p$ represent infinitely small number and infinitely large number respectively. 
Then, for each data $e$ in each file, it joins in the bucket buffer $\mathcal{B}_i$ since $e$ is bigger than $P_{i-1}$  and smaller or equal than $P_i$ (lines 11-13).  
If the bucket buffer $\mathcal{B}_i$ is full, then all data in $\mathcal{B}_i$ is written into the corresponding file (lines 14-16).

After each pass, the original data file is removed and many bucket files are generated (line 18). 
In next pass, it handles the data file generated by last pass.

After finishing all passes, the approximate sorting is generated by all remained bucket files (line 23). Since each bucket file will generate at most $p$ buckets in next pass.
Thus, the number of bucket file is at most $p^k$ when algorithm stops.

It is worth noting that picking random $\frac{m}{b}$ blocks  is the same as reading first $\frac{m}{b}$ blocks into main memory.
Because, we assume that the input permutation is uniformly random chosen from the permutation space $S_n$. 
Thus, the position of all data items are independent.
After seeing a certain amount of items sequentially, considering the next position, all data items unseen will have equal probabilities to be accessed.
Thus, simply reading the first $\frac{m}{b}$ blocks is the same as randomly picking $\frac{m}{b}$ items. 

\subsection{Analysis }
\subsubsection{Average ESP metric of Approximate Results}
Next, we analyze the average ESP metric of approximate results generated by Algorithm \ref{alg:1}. Before that, we give a conclusion about ESP metric.
\begin{lemma}
	\label{lemma:average esp}
	Let $\pi$ and $\delta$ be a permutation chosen independently and uniformly in $S_n$, Then
	\begin{equation*}
		E[d_{esp}(\pi,\delta)] = \frac{1}{3}\cdot \frac{n^2}{b}-O(b).
	\end{equation*}
\end{lemma}

	\begin{proof}
		According to the right invariant of ESP metric, the distribution of $d(\pi,\delta)$ is equal to the $d_{esp}(id,\pi)=d_{esp}(\pi)$.
		So we have
		\begin{equation*}
			E[d_{esp}(\pi)] = \sum_{i=1}^n {\frac{1}{n} \sum_{j=1}^n\left|\lceil\frac{i}{b}\rceil-\lceil\frac{j}{b}\rceil\right|} = \frac{(2b^2)}{n}\sum_{i=1}^{\frac{n}{b}-1}(1+i)\cdot i=\frac{1}{3}\frac{n^2}{b}-O(b)
		\end{equation*}
		\qed\end{proof}

This lemma shows that the average ESP metric among all permutations is $\frac{1}{3}\cdot \frac{n^2}{b}-O(b)$. 
The result of Algorithm \ref{alg:1} is composed by $p$ buckets. 
Only elements in same buckets are in wrong order. 
Thus, if we could get the average length of each buckets, the average ESP metric of whole sequence can be obtained.

\begin{theorem}
	\label{theorem: esp of al}
	Suppose that the input permutation $X$ of Algorithm 1 is randomly and independently chosen from $S_n$ and produces an output $Y$ with average ESP distortion $r$. We have
	\begin{equation}
		E[d_{esp}(Y,id)] < O(\frac{n^2}{m^k}\cdot b^{k-1}) + O(\frac{n^2}{m^{2k}}\cdot b^{2k}) -O(1).
	\end{equation}
\end{theorem}

	\begin{proof}
		In each pass, Algorithm \ref{alg:1} is get first $m$ data items  as samples to get $p-1$ pivots. 
		Thus, after $k$ passes, the number of samples is $p^{k-1}m$ and the number of buckets is $p^k$.
		The number of possible case of these $p^{k-1}m$ elements is $\tbinom{n}{p^{k-1}m}$. 
		
		From Lemma \ref{lemma:average esp}, it shows the average ESP metric of a random permutation of length $n$. 
		Let the $l_i,(1\leq i \leq p^k)$, as the length of buckets $i$ obtained by algorithm. 
		Hence, we obtain
		\begin{equation*}
			E[d_{esp}(Y,id)]=\tbinom{n}{p^{k-1}m}^{-1}\sum_{possible\,samples}\sum_{1\leq i \leq p^k}{\frac{1}{3}\frac{l_i^2}{b}-o(b)}
		\end{equation*}
		
		To compute the above sum, we count how many times a bucket of length $l$ happens for all possible values of samples.
		We know that the size of each bucket at least is $\lfloor \frac{m}{p} \rfloor$. 
		Without loss of generality, we assume that $p$ can divide $m$. 
		Thus, $\frac{m}{p}\leq l \leq n-p^{k-1}m+\frac{m}{p}$. 
		For bucket $\mathcal{B}_i (1<i<p^k)$ and its length is $l$,  its possible start position $j$ satisfies $(i-1)\cdot \frac{m}{p}+1 \leq j \leq n-l-p^{k-1}m+i\cdot \frac{m}{p}$ and end position is $j+l$. 
		Then, the number of buckets with length $l$ appearing starting at position $j$ equals $\tbinom{j-2}{(i-1)\cdot \frac{m}{p}-1}\tbinom{l-1}{\frac{m}{p}-1}\tbinom{n-j-l}{p^{k-1}m-i\cdot \frac{m}{p}}$.  
		For first and last bucket $\mathcal{B}_1$ and $\mathcal{B}_{p^k}$, the beginning point of $\mathcal{B}_1$ and the end position of $\mathcal{B}_{p^k}$ is determined. 
		Thus, for each $l$, the number of $\mathcal{B}_1$ with length $l$ is $\tbinom{l-1}{\frac{m}{p}-1}\tbinom{n-l}{p^{k-1}m-\frac{m}{p}}$ and the number of $\mathcal{B}_p$ with length $l$  is $\tbinom{l}{\frac{m}{p}}\tbinom{n-l-1}{p^{k-1}-\frac{m}{p}-1}$. 
		Hence, we have
		
		\begin{align*}
			E[d(Y,id)]=\tbinom{n}{p^{k-1}m}^{-1}\sum_{l=\frac{m}{p}}^{n-p^{k-1}m+\frac{m}{p}}\left(\frac{1}{3}\frac{l^2}{b}-o(b)\right)\cdot freq;
		\end{align*}  
		where 
		\begin{align*}
			freq = \tbinom{l-1}{\frac{m}{p}-1}\tbinom{n-l}{p^{k-1}m-\frac{m}{p}} &+ \tbinom{l}{\frac{m}{p}}\tbinom{n-l-1}{p^{k-1}m-\frac{m}{p}-1} \\ &+\sum_{i=2}^{p^k-1}\sum_{j=(i-1)\cdot\frac{m}{p}+1}^{n-l-p^{k-1}m+i\cdot\frac{m}{p}}\tbinom{j-2}{(i-1)\cdot \frac{m}{p}-1}\tbinom{l-1}{\frac{m}{p}-1}\tbinom{n-j-l}{p^{k-1}m-i\cdot \frac{m}{p}};
		\end{align*}
		Simplify above formula, we can get
		\begin{equation*}
			freq = \tbinom{l-1}{\frac{m}{p}-1}\tbinom{n-l}{p^{k-1}m-\frac{m}{p}} + \tbinom{l}{\frac{m}{p}}\tbinom{n-l-1}{p^{k-1}m-\frac{m}{p}-1} +(p^{k}-2)\tbinom{l-1}{\frac{m}{p}-1}\tbinom{n-l-1}{p^{k-1}m-\frac{m}{p}};
		\end{equation*}
		Then, it is easy to get that
		
		\begin{align*}
			E[d(Y,id)] &< O(\frac{n^2}{m^k}\cdot b^{k-1}) + O(\frac{n^2}{m^{2k}}\cdot b^{2k}) -O(1)	
		\end{align*}
		\qed\end{proof}

\begin{corollary}
	In I/O mode, Algorithm 1 is asymptotically optimal on ESP metric with a constant factor.
\end{corollary}

	\begin{proof}
		From upper bound of $r$ in Theorem \ref{theorem: esp of al}, we can know $r<O(\frac{n^2}{m^k}\cdot b^{k-1}) + O(\frac{n^2}{m^{2k}}\cdot b^{2k}) -O(1)	$. 
		From Theorem \ref{theorem:esp}, we know that the optimal algorithm with ESP metric $r^*\geq  O\left( \frac{n^2b^{k-2}}{ (em)^{k}}\right)$.
		Hence, we have
		\begin{equation*}
			\frac{r}{r^*} < O(b)
		\end{equation*}        
		Where $b$ is the number of data items contained in an I/O block.  
		$b$ is a constant is a constant factor related to practical application. 
		Thus, Algorithm 1 is asymptotically optimal with a constant factor.
		\qed\end{proof}

\subsubsection{Average EE metric of Approximate Results }
Next, we analyze the average ESP metric of approximate results of Algorithm \ref{alg:1}. Before that, we also present a conclusion about ESP metric.
\begin{lemma}
	\label{lemma:average ee}
	Let $\pi$ and $\delta$ be a permutation chosen independently and uniformly in $S_n$, Then
	\begin{equation*}
		E[d_{ee}(\pi,\delta)] = n-b.
	\end{equation*}
\end{lemma}

	\begin{proof}
		According to the right invariant of EE metric, the distribution of $d(\pi,\delta)$ is equal to the $d(id,\pi)=d(\pi)$.
		So we have
		\begin{equation*}
			E[d(\pi)] = \sum_{i=1}^n {\frac{1}{n} \sum_{j=1}^n\lceil\frac{i}{b}\rceil\oplus\lceil\frac{j}{b}\rceil} = n-b.
		\end{equation*}
		\qed\end{proof}

\begin{theorem}
	\label{theorem:ee of al}
	Suppose that the input permutation $X$ of Algorithm 1 is randomly and independently chosen from $S_n$ and produces an output $Y$ with average EE metric $r$. We have
	\begin{equation}
		E[d_{ee}(Y,id)] < (1-\frac{m^k}{nb^{k-1}})(n-\frac{m^k}{b^{k-1}}) +2n\frac{b^k}{m^k}.
	\end{equation}
\end{theorem}

	\begin{proof}
		Similar to the proof of Theorem \ref{theorem: esp of al}, 
		\begin{align*}
			E[d(Y,id)]=\tbinom{n}{p^{k-1}m}^{-1}\sum_{l=\frac{m}{p}}^{n-p^{k-1}m+\frac{m}{p}}\left(l-b\right)\cdot freq;
		\end{align*}
		where 
		\begin{equation*}
			freq = \tbinom{l-1}{\frac{m}{p}-1}\tbinom{n-l}{p^{k-1}m-\frac{m}{p}} + \tbinom{l}{\frac{m}{p}}\tbinom{n-l-1}{p^{k-1}m-\frac{m}{p}-1} +(p^{k}-2)\tbinom{l-1}{\frac{m}{p}-1}\tbinom{n-l-1}{p^{k-1}m-\frac{m}{p}};
		\end{equation*}
		Then, it is easy to get that 
		\begin{equation*}
			E[d_{ee}(Y,id)] < (1-\frac{m^k}{nb^{k-1}})(n-\frac{m^k}{b^{k-1}}) + 2n\frac{b^k}{m^k}.
		\end{equation*}
		\qed\end{proof}

\begin{corollary}
	In I/O mode, Algorithm \ref{alg:1} is asymptotically optimal on EE metric.
\end{corollary}

	\begin{proof}
		From Theorem \ref{theorem:external error}, $\alpha \geq 1-\frac{k}{c}$ after $k$ passes I/O where $\frac{n}{b} = O(\frac{m}{b})^c$.
		From Theorem \ref{theorem:ee of al}, $alpha <  (1-\frac{m^k}{nb^{k-1}})^2+2\frac{b^k}{m^k}$.
		It is easy to get that $(1-\frac{m^k}{nb^{k-1}})^2$ is slightly large than $1-\frac{k}{c}$.
		Thus, Algorithm \ref{alg:1} is asymptotically optimal for $k>0$. 
		\qed\end{proof}

\section{Applications on Approximate Sorting Results}
\label{sec:apps}
In this section, we analyze the applications on the data sequence or relations generated by our approximate sorting algorithm.
In fact, considering applications on approximate results is very difficult for mainly two reasons.

On the one hand, unordered sequence is a chaos for most of applications.
For example, it may causes great impact on the query result only if one element is out of order.
The binary search may cause mistakes on approximate results.
Disser et al.\cite{Disser2017} shows a search strategy to find the item using at least needs $\log \frac{n}{d}+2*d+O(1)$ queries, where $d$ is Spearman's footrule metric.
According to the analysis in Section \ref{sec:preliminary}, $d$ is large than $n$ in most case.
Thus, the worst case of searching an element in approximate result is to scan all items.

On the other hand, there are many possible permutations of approximate results with the same distortion metric.
For example, both ESP metric of the sequence $(8,2,3,4,5,6,7,1)$ and the sequence $(3,2,5,4,1,7,6,8)$ are $6$ for $b = 2$.
Thus, this differences between approximate results also leads to the different strategies for applications.

Thus, the approximate sorting result needs to be normalized, and make it possible to used by other applications.
That is also the function of the approximate sorting algorithm.
For our k-pass I/O approximate algorithm, the result sequence is globally ordered and locally unordered.
It divides the result sequence into many buckets, which data items are ordered among buckets and unordered in each bucket.

Based on that, we provide an index on this approximate result and the single and range query complexity with this index. 
Further, we also analysis the sort-merge join on our approximate results.

\subsection{Index on Approximate Results}
\label{subsec:app-index}
It is well know that data stored on the secondary storage of the system can be accessed either by a sequential scan or by using an index for randomly searches.
There are many indexes proposed in modern computer system.
The most common types of indexes are $B^+$-Trees and hash indexes.
$B^+$-Tree offers logarithmic, to the number of the data items, number of random accesses and support ordered range scans, which is wildly used.

\begin{figure}[h]
	\label{fig:index}
	\centering
	\subfigure[BF-Tree]{
		\label{fig:bf-tree}
		\includegraphics[width=0.47\textwidth]{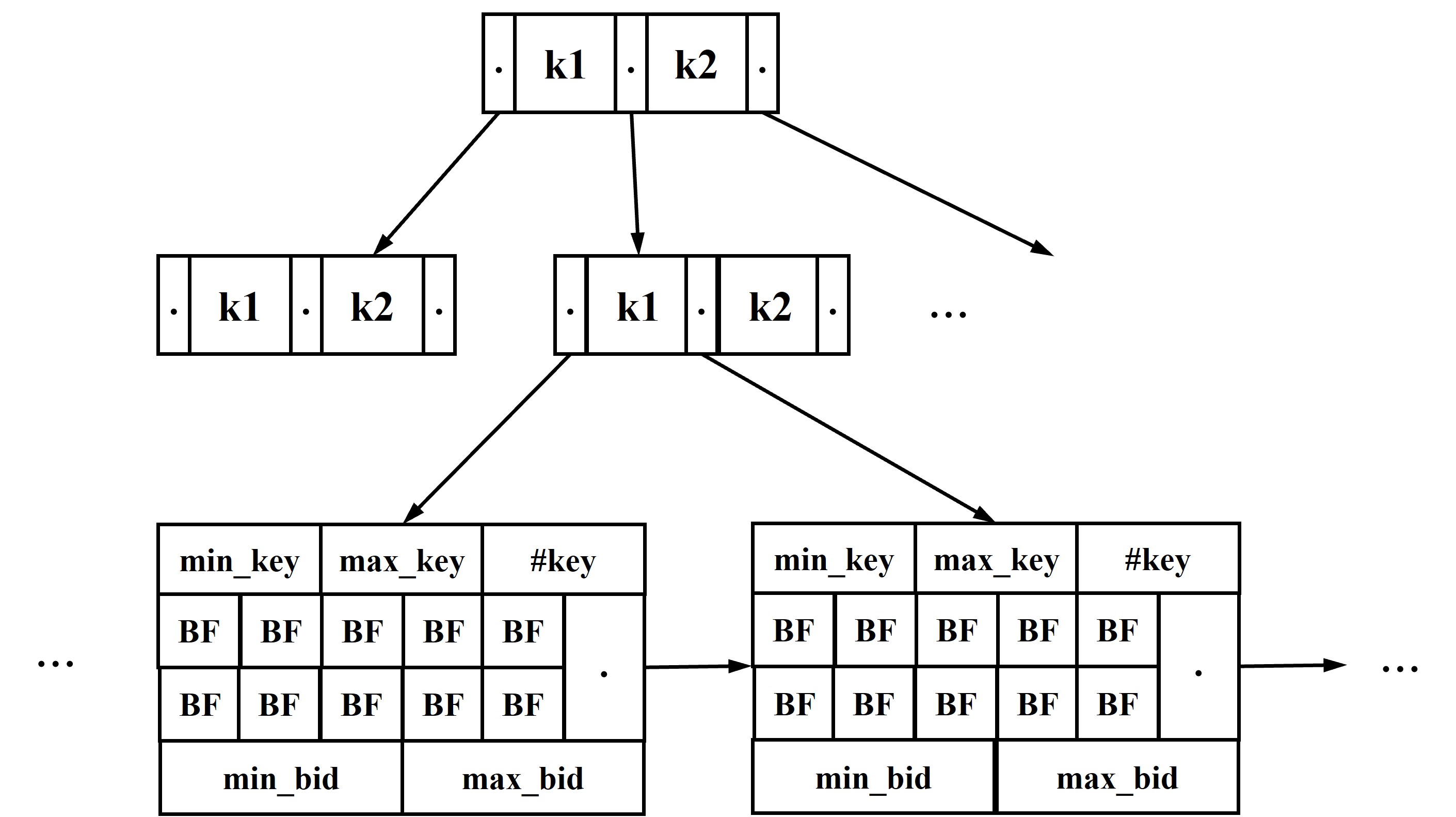}
	}
	\subfigure[VBF-Tree]{
		\label{fig:vbf-tree}
		\includegraphics[width=0.47\textwidth]{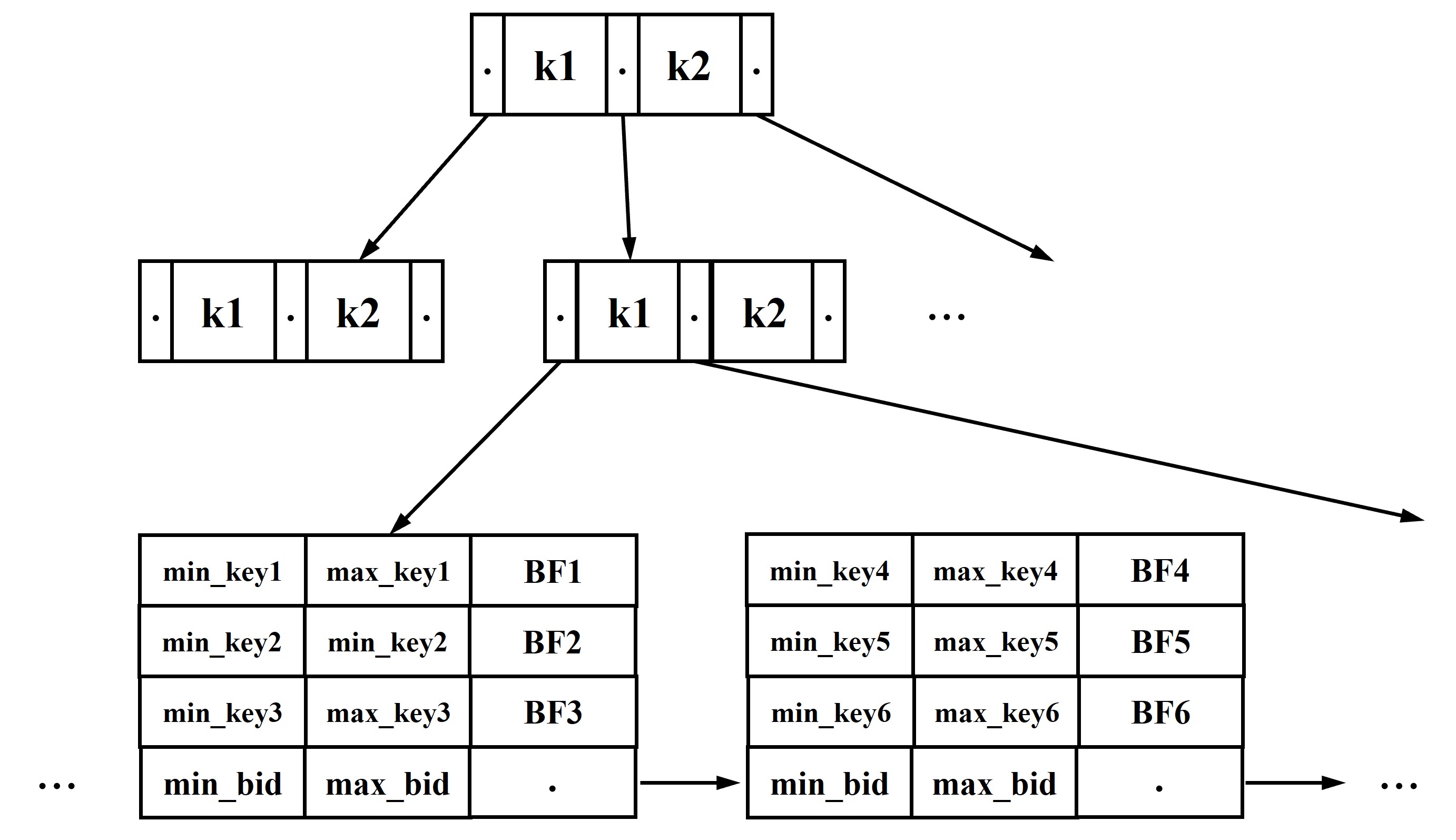}
	}
	\caption{The structure of tree index on approximate result}
	
\end{figure}

Manos et al.\cite{Athanassoulis2014} propose a tree structure index, named as Bloom Filter Tree (short for BF-Tree) , based on $B^+$-Tree, is to trade search accuracy for size and produce smaller.
BF-Tree also exploits the pre-existing ordering of the data to offer competitive search performance.
A BF-tree consists of nodes of two different types. The root and the internal nodes is the same as that in $B^+$-Tree.
It contains a list of keys with pointers between each pair of keys to other nodes in next level. 
Each leaf node of BF-Tree consists a key range $(min_key, max_key)$, an I/O block range $(min_bid, max_bid)$ and a number of Bloom Filters\cite{Bloom1970}, each of which stores the key for each block.
The leaf node also contains a pointer to the next leaf node.
The search operation on BF-tree is to retrieve the desired leaf node of BF-Tree, and performs a value probe for every Bloom Filter.
The probe result decides whether the key we search for exists in each Bloom Filter, with probability for false positive answer.
The structure of a BF-tree is shown in Figure \ref{fig:bf-tree}.

Our approximate sorting sequence is divided into many buckets by selected pivots. Data are ordered among buckets and unordered in each bucket.
In each bucket, the data are organized together in I/O blocks, and the data in each I/O block is sorted. 
Thus, it is obvious that the BF-tree is suitable for our approximate sorting results.
The internal nodes of BF-Tree are used to find the buckets where the data is located and the leaf nodes stores the Bloom Filter of each block in the bucket. 
During a BF-Tree index probe, it reads the BF-leaf which corresponds to the key and then probe all Bloom Filters.

In this paper, we provide a variant of BF-Tree, named as variant BF-tree (short for VBF-Tree), which is shown in Figure \ref{fig:vbf-tree}.
The root node and internal nodes is the same as that in BF-tree.
The main difference between BF-Tree and VBF-Tree is the structure of leaf nodes.

The VBF-Tree makes full use of the fact that data is ordered in each block of the bucket.
Rather than storing the a key range of the bucket, the leaf node in VBF-Tree contains the key range of each block in the bucket. 
Compare with the BF-Tree, the advantage of VBF-Tree is that it only needs to probe the Bloom Filters which the searched key is falling into the block range. 
If a Bloom Filter matches the searched key then its corresponding block contains the key with false positive probability $fpp$, and then such blocks will be read into internal memory.

In addition, the size of Bloom Filter is related to the number of elements it stores and the $fpp$.
When the number of elements $n$ the Bloom Filter stores is a constant, the $fpp$ is getting larger with the decrease of the size of Bloom Filter $s$. These three parameters of Bloom Filters is approximately satisfied by the formula\cite{Tarkoma2012}:
\begin{equation}
	\label{eq:bf}
	n = -\frac{s\cdot \ln^2 (2)}{\ln (fpp)} 
\end{equation}
Tree structures indexes are optimized for the external storage device, like HDD and SSD, and it sets the size of each node in indexes to match the I/O block size. 
When the data in each bucket is increasing,  the size of each Bloom filter will be decrease and the $fpp$ will increase.
Thus, to ensure a stable $fpp$, VBF-Tree allows the leaf node pointed by the pointer in upper level to split into multiple and connect by each other with the pointer in the leaf node.
The specific is shown in Figure \ref{fig:vbf-tree}.

\textbf{Creating a VBF-Tree:}
Due to that the VBF-Tree is a variation of BF-Tree for approximate sorting results generated by Algorithm \ref{alg:1}, the build time of VBF-Tree can be aggressively minimized using bulk loading, just like other tree indexes.
In order to bulk load a VBF-Tree, it creates the packed leaves of VBF-Tree firstly.
This can finish when Algorithm \ref{alg:1} outputs the approximate sorting results.
After that, it builds the remaining of the tree on top of the leaves level during a scanning of the leaves.

\begin{table*}[t]
	\centering
	\caption{The parameters of VBF-Tree }
	\label{table:para}
	\begin{tabular}{c|c}	
		\hline
		parameter name  & Description \\
		\hline
		$b$  & the number of keys in an I/O block \\
		$n$  & the number of elements \\
		$keysize$ & the size of the indexed key \\
		
		$ptrsize$ & the size of a pointer  \\
		
		$fpp$ & false positive probability   \\
		$fanout$ & fanout of the internal nodes in VBF-Tree \\
		$\#buckets$ & the number of the buckets \\
		$VBFh$ & the height of the VBF-Tree \\
		$BFsPerBlock$ & the number of BFs stored in an I/O block \\
		$leavesIO$ & the average number of leaves contained in each bucket\\
		$matchedBFs$ & the number of BFs whose range contains searched key\\
		$dataIO$ & the number of data blocks needed to be checked \\
		$\#bk$  & the number of buckets overlapped with the query range\\
		\hline 
	\end{tabular} 
\end{table*}

\subsection{Searching on Approximate Results}
\label{subsec:app-query}
In this subsection, we present an analytical model to VBF-Tree on the canonical approximate sorting results and 
calculate the I/O complexity for the single and range query on VBF-Tree.
Table \ref{table:para} shows the key parameters of the VBF-Tree.

\textbf{The cost of singe query:} The query on VBF-Tree is the same as it on BF-Tree. The cost of a singe query is comprised of three parts.

The first part is to retrieve the desired leaf node of VBF-Tree. Its I/O cost is equal to the height of VBF-Tree.
The $fanout$ of internal nodes of VBF-Tree is calculated  by  Equation \ref{eq:fanout}.
The number of buckets, $\#buckets$, is at most $p^k$, where $p$ is the number of buckets generated by Algorithm \ref{alg:1} in each pass.
The height of VBF-Tree is calculated by Equation \ref{eq:height}.
\begin{equation}
	\label{eq:fanout}
	fanout = \frac{b*keysize}{ptrsize+keysize}.
\end{equation}
\begin{equation}
	\label{eq:height}
	VBFh = \lceil\log_{fanout} \#buckets \rceil + 1 \leq k\cdot\lceil\log_{fanout} p \rceil + 1.
\end{equation}

The second part is to read all leaf nodes in the bucket into internal memory. Its I/O cost is the number of the leaf nodes in each bucket.
The number of Bloom Filters in a single leaf node is calculated by Equation \ref{eq:nbf}. As in Equation \ref{eq:bf}, the size of a single Bloom Filter is decide by $fpp$ and the number of elements it stores.
\begin{equation}
	\label{eq:nbf}
	BFsPerBlock = \frac{b*keysize-8-ptrsize}{2keysize-(\frac{b \cdot \ln(fpp)}{\ln^2(2)})}.
\end{equation}
The average number of blocks in each bucket is $\frac{n}{b\cdot p^k}$.
Thus, the average I/O cost for reading the leaf nodes is calculated by Equation \ref{eq:leavesIO}
\begin{equation}
	\label{eq:leavesIO}
	leavesIO = \frac{n}{b\cdot p^k \cdot BFsPerBlock}.
\end{equation}

The third part is to scan the corresponding leaf nodes and probe the searched key for each bloom filter whose key range contains the searched key. The I/O cost is equal to the matched bloom filter, which is shown in Equation \ref{eq:dataIO}.
\begin{equation}
	\label{eq:dataIO}
	dataIO = matchedBFs\cdot fpp
\end{equation}

Thus, the I/O cost of a single query on our approximate sorting result with VBF-Tree is $VBFh+leavesIO+dataIO$.

\textbf{The cost of range query:} The range query is much simpler than single query on VBF-Tree since it doesn't need to probe the Bloom Filter. 
The range query firstly finds the corresponding buckets of range boundary and check whether the range of each bloom filter is overlapped the query range.
If the range of each bloom filter is overlapped with the query range, then the corresponding data block is read into internal memory and the satisfying keys are found.
The cost of range query includes $2\cdot VBFh$ for finding  corresponding buckets.
Then, all the leaves in the buckets overlapped by query range need to be scanned.
We denote the number of  buckets overlapped by the query range is $\#bk$.
The average I/O cost for reading the leaves is $\#bk \cdot leavesIO$.
Finally, all the data blocks corresponding to the matched Bloom Filters are checked, and $dataIO$ becomes the number of $matchedBFs$.

Thus, the I/O cost of range query on our approximate sorting result with VBF-Tree is $2\cdot VBFh + \#bk \cdot leavesIO + matchedBFs$.

\subsection{Join on Approximate Results}
\label{subsec:app-join}
The join operation takes two input relations, $R_1$ and $R_2$, and produces a single resulting relation.
Many types of joins have been defined, including natural joins, semi-joins, equijoins, outer joins, etc.
The traditional sort-merge join algorithm maintains a read pointer for each sorted attribute of relation.
Then, it linearly scans two relations in an interleaved way, and will encounter these sets at the same time.
Furthermore, the result relation of sort-merge join is also sorted.

In this subsection, we discuss the sort-merge join algorithm on approximate sorting results generated by Algorithm \ref{alg:1}. The goals of  sort-merge join algorithm on approximate sorted results are to incur no disk overhead under low distortion and perform efficiently under high distortion.
There are two types of approximate sorted relations on join operation. 
\begin{enumerate}
	\item[1)]Only one side relation, $R_1$ or $R_2$, is approximate sorted, the other relation  is totally sorted.
	\item[2)]Both relations are approximate sorted.
\end{enumerate}

As illustrated in Section \ref{subsec:app-index}, our approximate sorted relation is ordered among buckets and unordered in each bucket. 
The main challenge for sort-merge join on the approximate sorted relations is finding the corresponding key in each unsorted bucket.
Firstly, we show a definition about the unsorted bucket.
\begin{definition}
	\label{def:m-tolerable}
	Denote $N_{B_i}$ as the number of data item in each bucket $B_i$. If $\frac{N_{B_i}}{b} \leq \frac{m}{b} - 2$, we call this bucket is $m$-$tolerable$.
\end{definition}

There are more than two fewer blocks in the $m$-$tolerable$ bucket than in internal memory.
These two blocks are reside for the other input relation and result.
Since all tuples in each bucket can be read into internal memory, we can use the internal sorting algorithm to make the unordered data in each bucket into ordered. 
Join operation on such $m$-$tolerable$ buckets is much easier.
Next, we analyze the sort-merge join for each type.

\begin{figure}[t]
	\label{fig:one-side}
	\centering
	\subfigure[case1]{
		\label{fig:one-side1}
		\includegraphics[width=0.40\textwidth]{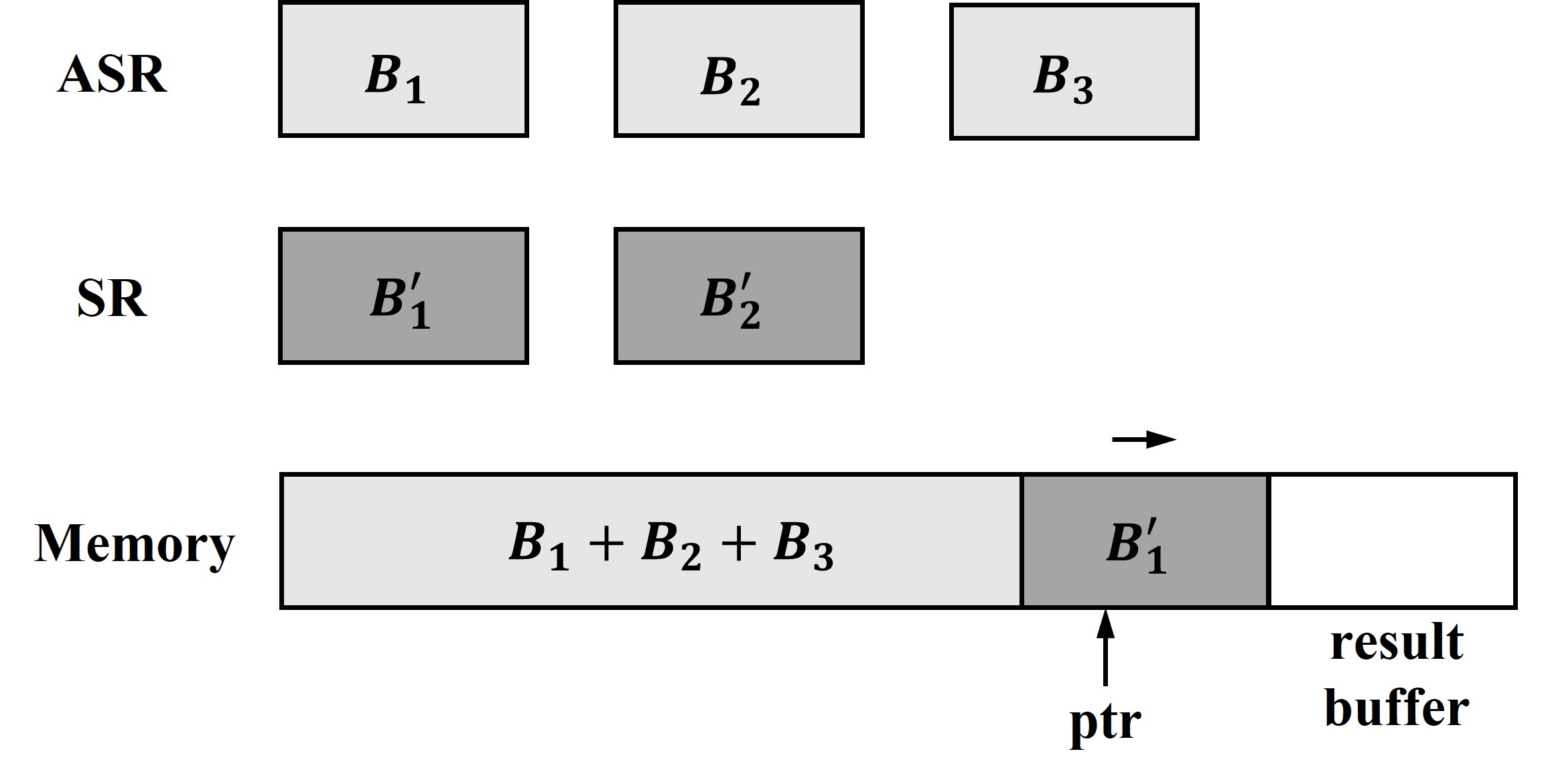}
	}
	\subfigure[case2]{
		\label{fig:one-side2}
		\includegraphics[width=0.40\textwidth]{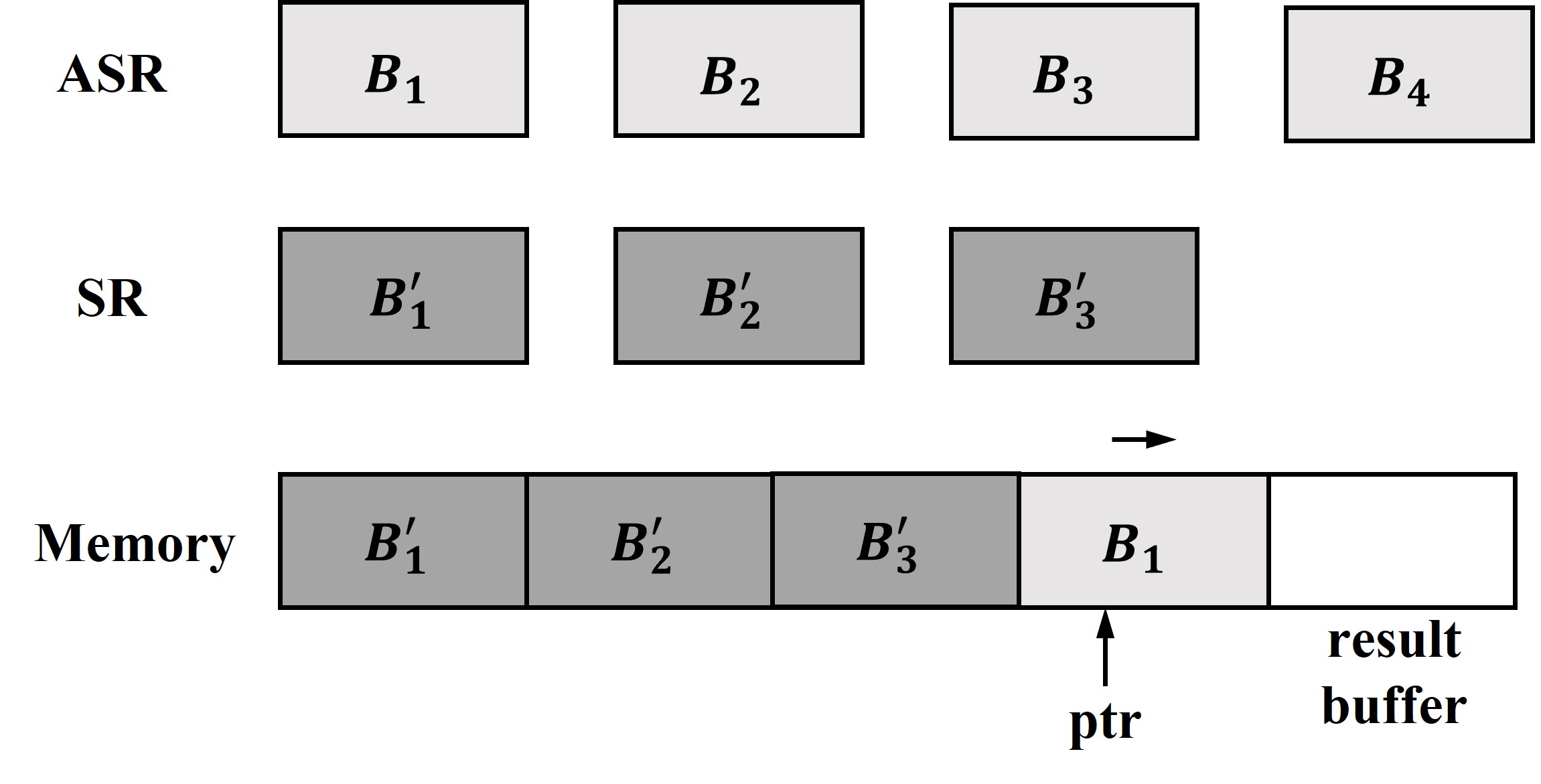}
	}
	\subfigure[case3]{
		\label{fig:one-side3}
		\includegraphics[width=0.40\textwidth]{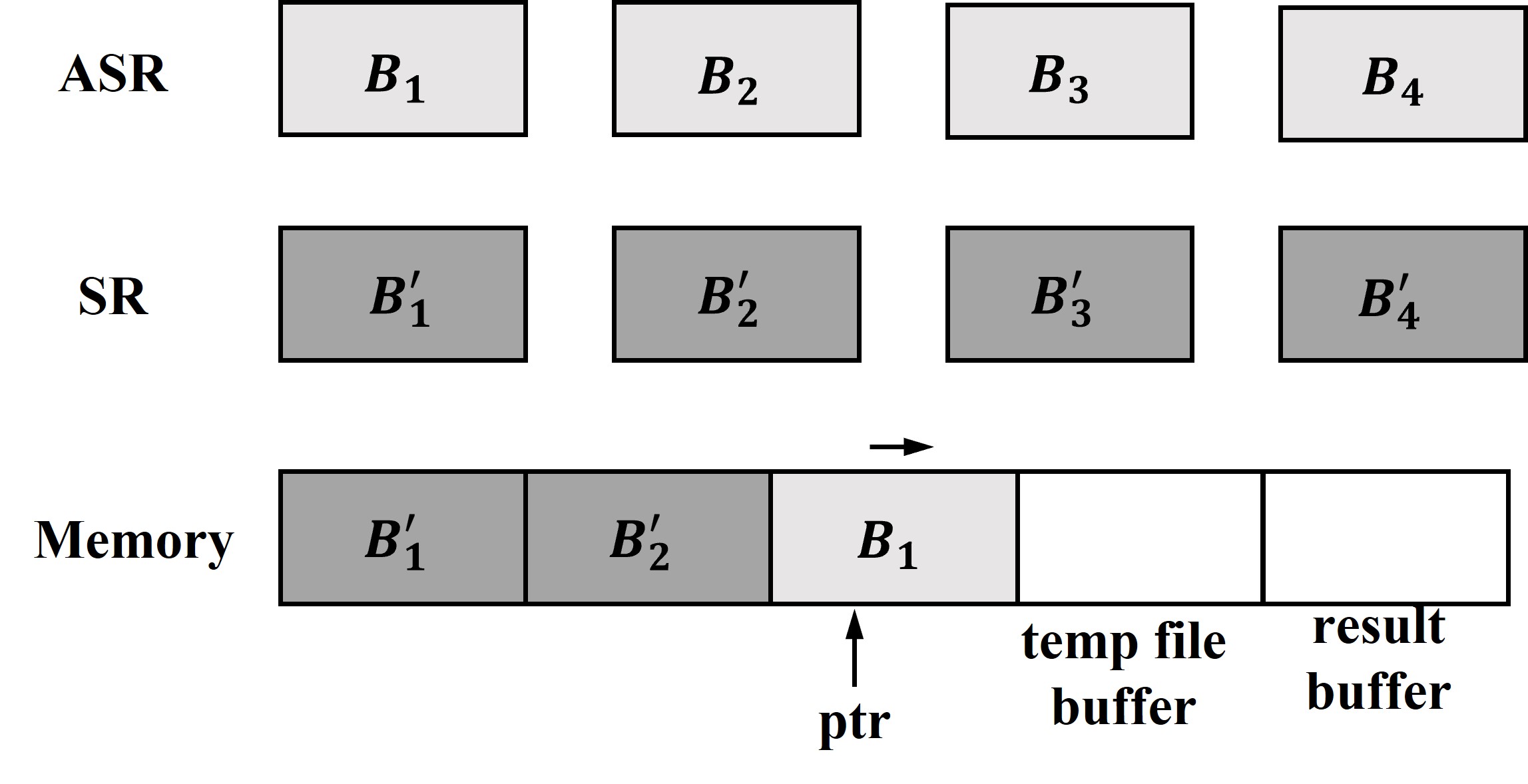}
	}
	\subfigure[another method]{
		\label{fig:one-side4}
		\includegraphics[width=0.40\textwidth]{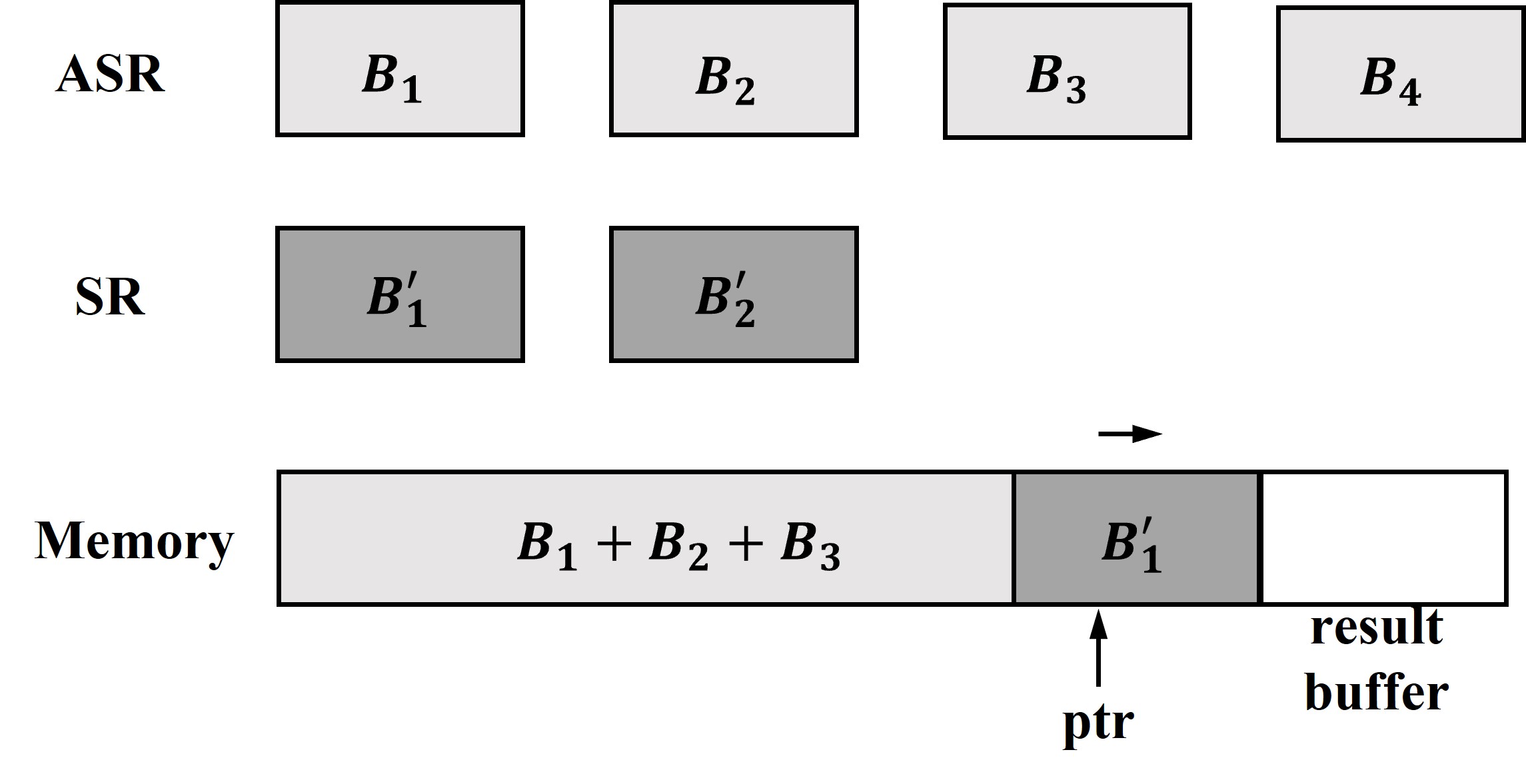}
	}
	\caption{The example for join while one-side relation is approximate sorted}
	
\end{figure}

\textbf{One-side relation is approximate sorted:} 
Figure \ref{fig:one-side} depicts the different cases for sort-merge join on one-side approximate sorted relations. In Figure \ref{fig:one-side}, ASR means the approximate sorted relation and SR means the sorted relation.
The data blocks $B_1$,$B_2$,$B_3$ and $B_4$ in Figure \ref{fig:one-side} are all in one bucket.

The \textbf{first case} is that the bucket of ASR is $m$-$tolerable$, which is shown in Figure \ref{fig:one-side1}. 
In this case, it only needs to read all the data blocks in the bucket of ASR  into the internal memory, and sort the data.
Next, it linearly scans the bucket data in memory and the data of SR in a interleaved way and outputs the matching results just like the traditional sort-merge join.

The \textbf{second case} is that the bucket of ASR is not $m$-$tolerable$, but the data in SR corresponding to key range of this bucket can be read into the internal memory, which is shown in Figure \ref{fig:one-side2}.
In this case, we set the key range of the bucket in ASR be $(K_{min}, K_{max})$.
If the data of SR which is in the key range $(K_{min}, K_{max})$ can all be read into internal memory, then all the data in the bucket of ASR can find their matched key in internal memory.
For example, if the data range of blocks $B'_1$, $B'_2$ and $B'_3$ in SR, is $(K'_{min}, K'_{max})$, which satisfies that $K'_{min} \leq K_{min}$ and $K'_{max} \geq K_{max}$, then it reads the blocks $B'_1$, $B'_2$ and $B'_3$ into memory.
Thus, all the data in the bucket of ASR will find the corresponding key in SR.
In this case, the I/O cost is also the same as the traditional sort-merge join.

The \textbf{third case} is that the bucket of ASR is not $m$-$tolerable$ and the size of data in SR corresponding to key range of this bucket is large than the internal memory, which is shown in Figure \ref{fig:one-side3}.
The key range of the bucket in ASR is $(K_{min}, K_{max})$, but the corresponding data blocks $B'_1$, $B'_2$, $B'_3$ and $B'_4$ in SR cannot be read into internal memory.
In this case, it first reserves a buffer for a temp file in internal memory and read as many data blocks in SR as possible.
For example, we set the key range of blocks $B'_1$, $B'_2$ be $(K'_{min}, K'_{mid})$, which satisfies that $K'_{min} \leq K_{min}$ and $K'_{mid} \leq K_{max}$. 
Thus, after a single scan of the data in the bucket of ASR, only the data in $(K_{min}, K'_{mid})$ find their matched keys in SR and the data in $(K'_{mid}, K_{max})$ need to be processed again.
Here, the temp file is used to record the data of ASR in  $(K'_{mid}, K_{max})$ to reduce the unnecessary duplicate processing for the data of ASR in $(K_{min}, K'_{mid})$.
In next step, the data blocks $B'_3$ and $B'_4$ will be read into internal memory and the data in temp file will be scanned.
By this way, all the data in SR are read sequentially, and only out-of-range data in ASR will be written and read repeatedly.

According to the description of these three cases, we notice that the solutions of sort-merge join in the first two cases are optimal, since the I/O cost of such cases is same as the I/O cost of sort-merge join on totally sorted relations.
In fact, if the bucket of ASR is not $m$-$tolerable$, it can also use the same method as in first case, which is shown in Figure \ref{fig:one-side4}.
By this way, it needs to re-scan the data in SR another time when data block $B_4$ is read into internal memory.
This method eliminates the impact of approximate order by scanning the data in the ordered relation multiple times.
However, the optimal solution in the second case will never be used.
Furthermore, the data needed to be re-scanned is gradually decreasing in the third case, and it won't by this way.

In addition, this type of sort-merge join can be used when one relation is relatively large than the other.
We can adopt the approximate sorting to generate an approximate sorted sequence for the larger relation, while the smaller relation is totally sorted. 
It may satisfies the second case with high probability for smaller relation, since the data in key range $(K_{min}, K_{max})$ of smaller relation is more likely to be read into internal memory. 

Since the bucket size in approximate sorted relation is not same, it is difficult to compute the 
I/O cost for sort-merge join.
For two relations ASR and SR, the quantity of data item is $n_1$ and $n_2$ respectively.
The best case I/O cost of sort-merge join is $O(\frac{n_1}{b} + \frac{n_2}{b} + k\cdot \frac{n_1}{b} + \frac{n_2}{b}\log(\frac{n_2}{b}))$ ($n_1>n_2$ and ASR is approximate sorted), at which all the buckets in approximate sorted relation are processed by the first case or second case.
The worst case I/O cost is $O(\frac{n_1}{b}\cdot \frac{n_2}{m\cdot p^k}+\frac{n_2}{b}+k\cdot \frac{n_1}{b} + \frac{n_2}{b}\log(\frac{n_2}{b}))$, at which all the buckets in approximate sorted relation are processed by the third case.

\begin{figure}[h]
	\label{fig:two-side}
	\centering
	\subfigure[case1]{
		\label{fig:two-side1}
		\includegraphics[width=0.40\textwidth]{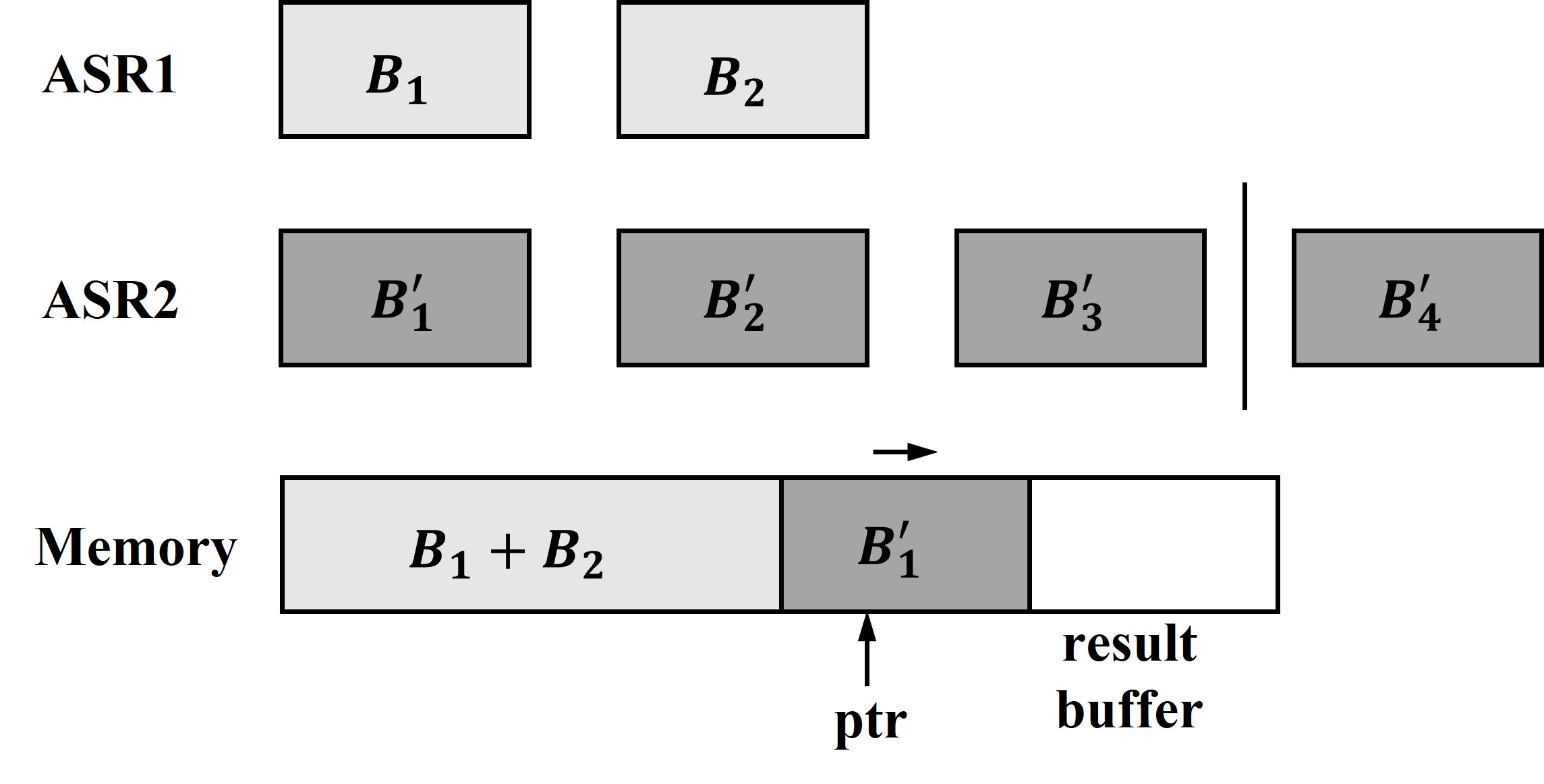}
	}
	\subfigure[case2]{
		\label{fig:two-side2}
		\includegraphics[width=0.40\textwidth]{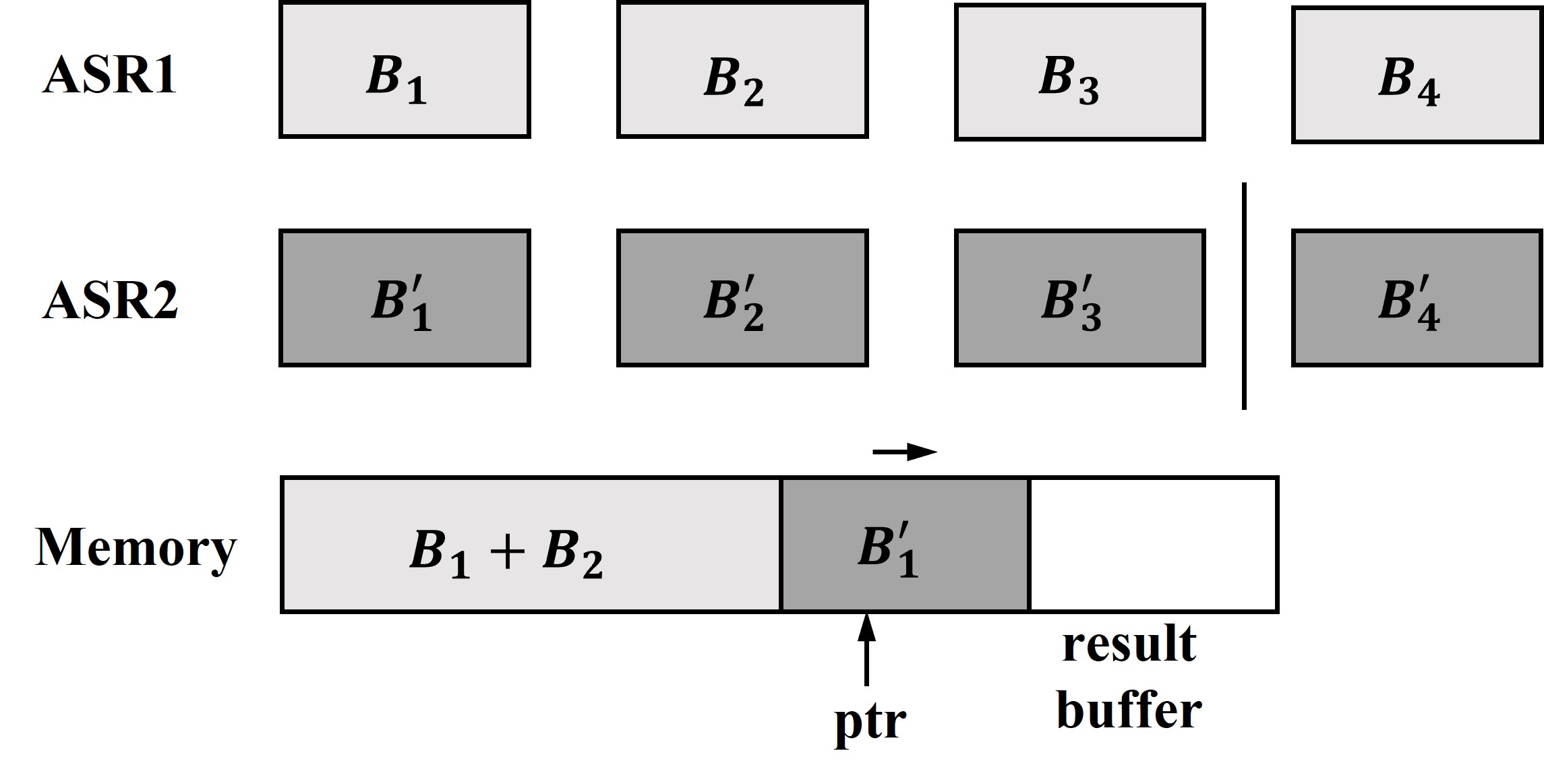}
	}
	\caption{The example for join while both relations are approximate sorted}
	
\end{figure}
\textbf{Both relations are approximate sorted:} Figure \ref{fig:two-side} shows the cases for join on two approximate sorted relations. 
Due to that the range of each bucket in two relation are not the same, there is  no extra optimization option for join on both approximate sorted relation, and it may needs to re-scan the related bucket multiple times to get the right join result.

Like in the one-side analysis, the \textbf{first case} is that a bucket in one of the two relations is $m$-$tolerable$, which is shown in Figure \ref{fig:two-side1}. 
In Figure \ref{fig:two-side1}, the bucket of ASR1 is read into memory.
The data blocks $B'_1$, $B'_2$ and $B'_3$ is in a bucket and block $B'_4$ is in another adjacent bucket.
We set the key range of bucket in ASR1 be $(K_{min}, K_{max})$.
Only if the key range of the data blocks $B'_1$, $B'_2$ and $B'_3$ in ASR2 is equal to the key range of bucket in ASR1,
the I/O cost of sort-merge join on approximate sorted relations is the same as the I/O cost  of sort-merge join on sorted data.
Whatever the key range of the data blocks is cross or overlapped the range  $(K_{min}, K_{max})$, it means that there exist the data in the bucket of ASR2 which is not matched with any data in the bucket of ASR1.
Thus, this bucket of ASR2 needs to be re-scanned for another bucket of ASR1.
For example, we set the key range of the bucket in ASR2 be $(K'_{min},K'_{max})$.
If $K'_{max} > K_{max}$, it means that the data blocks $B'_1$, $B'_2$ and $B'_3$ need to be scanned again for the data in ASR1 in $(K_{max},K'_{max})$ to find the corresponding key.
If $K'_{max} < K_{max}$, it needs to scan the data in the next bucket of ASR2 for the data in $B_1$ and $B_2$ of ASR1, and the bucket where data block $B'_4$ is located needs to be scanned.
In conclusion, for a bucket of ASR1, all the buckets in ASR2, whose key range is cross the $(K_{min},K_{max})$, are needed to be scanned multiple times when the key range of buckets in two relations are not same.

The \textbf{second case} is that none of the buckets in  two relations is $m$-$tolerable$, which is shown in Figure \ref{fig:two-side2}.
In this case, we choose to read the data block of large bucket of two relations into the internal memory.
In Figure \ref{fig:two-side2}, it first processes the data blocks $B_1$ and $B_2$ in ASR1 using the same way as in the first case.
After finishing the $B_1$ and $B_2$, the data blocks $B_3$ and $B_4$ are read into the memory and treated in the same way until all the block in the bucket of ASR1 is finishing.

In general, the sort-merge join on two approximate sorted relations is very inefficient unless the bucket range of ASR1 and ASR2 is same and a bucket in one of the two relations is $m$-$tolerable$.
It is also difficult to compute the I/O cost.
For two relations ASR1 and ASR2, the quantity of data item is $n_1$ and $n_2$ respectively.
The best case I/O cost is $O(\frac{n_1}{b} + \frac{n_2}{b} + k\cdot \frac{n_1}{b} + k\frac{n_2}{b})$, when all of the buckets in one of relations are $m$-$tolerable$ and the range of each bucket in two relation are same.
The worst case I/O cost is $O(\frac{n_1}{b} + \#bk\cdot\frac{n_2}{b} + k\cdot \frac{n_1}{b} + k\frac{n_2}{b})$, when all the buckets are not $m$-$tolerable$.
The $\#bk$ is the average number of  buckets in ASR2 crossed with range of each bucket in ASR1.

\section{Conclusion}\label{sec:conclusion}
In this paper, we analyze the approximate sorting in I/O model. To better measure the quality of approximate result in I/O model, we propose a new type of metric, which named \textit{External metric}. It ignores the errors and dislocations happened in each I/O block.
Specifically, we use \textit{External Spearman's footrule} metric as an example to analyze the approximate sorting in I/O model.
In addition, we propose a new metric \textit{external errors} to directly state the number of dislocation elements in approximate result.
Then, according to the rate-distortion relationship endowed by these two metrics, the lower bound of these two metrics on external approximate sorting problem with $t$ I/O operations is proved.
We propose a k-pass external approximate sorting algorithm, named as EASORT, and prove that EASORT is asymptotically optimal.
Finally, we consider the applications on approximate sorting results. An index for the result of our approximate sorting is proposed and analyze the single and range query on approximate sorted result using this index.
Further, the sort-merge join on two relations, where one of the relations is approximate sorted or both relations are approximate sorted, are all discussed in this paper.  

%
%
%
 \bibliographystyle{splncs04}
 \bibliography{approximatesorting.bib}

\end{document}